\documentclass{article}

\usepackage{arxiv}

\usepackage[utf8]{inputenc} % allow utf-8 input
\usepackage[T1]{fontenc}    % use 8-bit T1 fonts
\usepackage{hyperref}       % hyperlinks
\usepackage{url}            % simple URL typesetting
\usepackage{booktabs}       % professional-quality tables
\usepackage{amsfonts}       % blackboard math symbols
\usepackage{nicefrac}       % compact symbols for 1/2, etc.
\usepackage{microtype}      % microtypography
\usepackage{lipsum}		% Can be removed after putting your text content
\usepackage{graphicx}
\usepackage[square,sort,comma,numbers]{natbib}
\usepackage{doi}
\usepackage{dsfont}
\usepackage{amsmath}
\usepackage{tikz}
\usepackage{subfigure}
\usepackage{diagbox}
\usetikzlibrary{positioning}
\usetikzlibrary{arrows.meta,arrows}

\title{Learning Chemotherapy Drug Action via Universal
Physics-Informed Neural Networks}

\author{Lena Podina \\
	Cheriton School of Computer Science\\
	University of Waterloo\\
	Waterloo, ON, Canada\\
	\texttt{lpodina@uwaterloo.ca} \\
	\And
	Ali Ghodsi \\
	Cheriton School of Computer Science\\
	University of Waterloo\\
	Waterloo, ON, Canada \\
	\texttt{ali.ghodsi@uwaterloo.ca} \\
	\AND
	Mohammad Kohandel \\
	Department of Applied Mathematics \\
	University of Waterloo\\
	Waterloo, ON, Canada\\
	\texttt{kohandel@uwaterloo.ca} \\
}

\renewcommand{\shorttitle}{Learning chemotherapy drug action via UPINNs}

\hypersetup{
pdftitle={Learning Chemotherapy Drug Action via Universal
Physics-Informed Neural Networks},
pdfsubject={q-bio.NC, q-bio.QM},
pdfauthor={David S.~Hippocampus, Elias D.~Striatum},
pdfkeywords={First keyword, Second keyword, More},
}

\begin{document}
\maketitle

% \begin{abstract}
% 	\lipsum[1]
% \end{abstract}

% keywords can be removed
% \keywords{First keyword \and Second keyword \and More}

\section{Introduction}\label{sec:intro}

Quantitative systems pharmacology (QSP) is a commonly used approach of mathematically assessing drug pharmacokinetics and pharmacodynamics before they go to clinical trial~\cite{wang2019model}. Many drugs fail phase II and III clinical trials~\cite{wong2019estimation} due to inadequate understanding of the mechanism of action. However, via QSP models, which integrate biological, physiological and pharmacological data~\cite{sorger2011quantitative}, it is possible to understand the underlying biological mechanisms and optimize drug dose and schedule. Additionally, QSP models can be used to predict drug toxicity or efficacy~\cite{polak2019better}, as well as individual patient response to the drug.

Despite its benefits, QSP suffers from several limitations which make the models costly and time-consuming to develop. First, many QSP models are built by making simplifying assumptions about the underlying biology. In general, manual distillation of the literature (and potentially large amounts of data) is done in order to build these assumptions accurately, rather than building them directly or automatically in a data-driven way. Secondly, building a model fully (including parameter estimation) can be very time-consuming due to the model complexity and manual distillation of the literature~\cite{twoheads}. QSP modelling is especially affected by these issues because a QSP model may comprise more than a dozen dynamical variables, and even more parameters~\cite{twoheads}. Recently, machine learning (ML) has been used to improve and simplify the QSP model-building process. Decision trees, regression, and neural networks are some of the tools that have been used to predict drug toxicity and patient response in a data-driven way~\cite{camacho2018next}. A major benefit of ML is the almost entirely automatic learning of patterns in the data by a model, and potentially robust generalization to unseen data~\cite{camacho2018next}. However, purely ML-based approaches often need large amounts of data to make accurate predictions, and are often uninterpretable. This means that although the model may produce reasonable predictions, the mechanism by which it does so is opaque. For this reason, an ML model may be sometimes called a ``black box''. In systems pharmacology, it is often important to know which features are the most important for prediction and why a certain outcome is expected. These findings shed light on underlying biological mechanisms, and inform drug development. This is also key to a sanity-check of the model prediction, and trust in the output.

Integrating ML and QSP is a promising research direction which has promise in binding together their strengths while compensating for their respective weaknesses. Machine learning can be used to automate the learning of complex patterns, or abstract away the function of less important mechanisms. QSP, on the other hand, can provide structure to the machine learning model and inform it with prior knowledge. An integration of these two methods means that some assumptions will be built in or enforced, but other interactions and mechanisms will be learned directly from the data. Due to the additional structure imposed by assumptions, potentially less data will be needed to learn the unknown components. The components of the model that need potentially more manual distillation can be learned more automatically from the data. Finally, the interpretability of the ML components may be augmented due to the surrounding structure imposed by the QSP model.

Existing integrations of QSP and ML tackle problems such as parameter inference~\cite{RAISSI2019686}, inference of model structure~\cite{eduati2020patient,rackauckas2020universal}, dimension reduction~\cite{derbalah2022framework}, and creation of virtual patients~\cite{allen2016efficient}. A prominent method to identify parameters from data is physics-informed neural networks (PINNs). PINNs are a flexible framework and have been extended in various ways~\cite{przedborski2021systems,yang2021b,jagtap2020extended} for the purposes of model fitting and inference, which are key problems in systems pharmacology. Given a differential equation that models a particular biological process, PINNs use a neural network to fit a surrogate solution. The unknown parameters are treated as additional weights to be optimized. However, PINNs cannot be used to identify entire unknown components of the model. Inference of model structure, as described in~\cite{twoheads}, has followed several different directions, such as logic-based modeling~\cite{eduati2020patient}, integration of QSP with genome-scale computational models~\cite{puniya2021integrative}, and universal differential equations (UDEs)~\cite{rackauckas2020universal}. UDE's are a popular method to identify unknown components of differential equations. However, they are not robust to noise, as~\cite{podina2022pinn} recently showed, and the Universal Physics-Informed Neural Network (UPINN) method is more robust. In this paper, we apply the UPINN method to chemotherapy modelling, as a proof of concept of what the method can accomplish in the realm of pharmacokinetic and pharmacodynamic modelling. We have applied the method to both simulated and in-vitro data, and showed high accuracy of identification of the unknown components of differential equations.

In the remaining sections, we start with an overview of existing modeling approaches to chemotherapy modelling, physics-informed neural networks and QSP structure identification methods. Following that, we use the UPINN method to identify the hidden components of various ODEs that model chemotherapy dynamics. We show the performance of our method on both synthetic and in-vitro data. For the experiments on in-vitro data, we learn the time-dependent effect of a chemotherapeutic, doxorubicin, on cell growth (proliferation).

Our contributions in this work are:
\begin{itemize}
    \item We integrate machine learning in the form of PINNs with QSP models in order to identify unknown drug actions within QSP models. To illustrate, we apply the Universal PINN method to identify hidden terms in QSP models in cases of both synthetic and in-vitro experimental data.
    \item Via simulations, we show that three different types of drug action (Log-Kill, Norton-Simon, $E_{max}$) can be identified from the chemotherapy concentration and number of cells over time. We also employ our approach to recover dose-dependent parameters from several sets of data simultaneously and interpolate these parameters between dosages. This could potentially replace the repeated application of a standard physics-informed neural network. % simulated data? Real data? Combine PINNs with UPINNs and infer parameters that way?
    % \item 
    \item We show that our approach can successfully identify the time-dependent net proliferation rate in cases of both synthetic and in-vitro experimental doxorubicin data.
\end{itemize}
% However, It is key to note that all of the PINN, UDE, and UPINN methods, do not require large amounts of data to be accurate in identifying the parameter or component. This showcases a key advantage to the integration of QSP and ML.
% Talk about alternatives to the UDE thing. Papers that cited it.
% Limitations of QSP... and how they can be addressed using ML
% Cite existing work

% ML is not without its downsides, and should be appropriately combined with QSP for maximal interpretability
% Sparse data regimes and how they can be combined well in these regimes
% Knowledge-guided machine learning

% Chemotherapy modelling: what has been done and ML for chemotherapy modelling
% We are able to bypass some of the assumptions.
% ML for chemotherapy modelling: RL?? etc.

% Contributions

% Explain what you did and why it's important

\section{Background}\label{sec:background}

% Talk about physics-informed neural networks

\subsection{Physics-informed Neural Networks}

Physics-informed neural networks~\cite{RAISSI2019686} (PINNs) were developed by Raissi et al. for the purposes of solving forward and inverse problems of differential equations (partial (PDE) and ordinary (ODE)). For the forward problem, a neural network is used to provide a numerical solution to a PDE using data that satisfies the PDE. In this case, the data generating process is fully known. For the inverse problem, the differential equation is known except for certain parameters, which take on a real value. Using data, PINNs can be used to identify the parameters that best fit the data.

As mentioned previously, a drawback of PINNs is that they cannot be used to identify entire hidden terms of differential equations. In~\cite{podina2022pinn}, a method called Universal PINNs (UPINNS) was developed to find the hidden terms in a data-driven way.  was demonstrated on the Lotka-Volterra system of ODEs, on a model of cell apoptosis with bistability, and on Burger's equation, a PDE. In this paper, we explore further the power of the method in identifying unknown terms from data, thereby avoiding modelling assumptions. We base our work on the models in~\cite{panetta2003optimal} and~\cite{kohandel2006mathematical} We show that in every combination of drug action, we are able to make an approximation of the drug action using a neural network. Furthermore, we can combine the UPINNs approach from~\cite{podina2022pinn} with the traditional PINN approach to identify parameters first, and then use these parameters to inform the identification of the hidden terms.

% add another formula to explain...

\subsection{Chemotherapy modelling}

Much of the effort of the modelling of chemotherapeutics has gone into optimizing the drug schedule. This means finding the dosing schedule that kills the most cancer cells, while limiting the toxicity to normal cells. In many studies, the modelling of the drug is performed using assumptions on how the drug acts on the cells, and afterwards, the parameters of the model may be fit from data. For example, in~\cite{traina2010optimizing}, a Norton-Simon model of drug action and Gompertzian growth was assumed, and then fit to data. After, the drug schedule can be perturbed and the best dosing schedule can be identified from model predictions. In~\cite{drexler2020experimental}, a reaction kinetics analogy is assumed for the model, and the parameters are fit to tumours in untreated mice. Fits are subsequently evaluated for treated mice.~\cite{basse2004modelling} used assumptions based on paclitaxel's effect on cells in different phases of the cell cycle. Hence their model is compartment based, with differential equations governing the transition of cells from one phase to the next.
In~\cite{jarrett2019experimentally}, tumour growth was modelled with logistic growth, but the effect of the chemotherapeutic (also paclitaxel) is modelled only through its effect on the carrying capacity.
In~\cite{bolton2015proposed}, a novel fractional-order Gompertz model is introduced and determined to be a better fit for experimental mouse tumour data. 

Optimizing the order and sequence of chemotherapy and surgery is also an important problem. In~\cite{kohandel2006mathematical}, a mathematical model of ovarian cancer was developed to determine whether chemotherapy, then surgery, or whether surgery then chemotherapy was a better treatment. The authors examined Gompertzian and logistic growth, and three different types of cell-kill methods: log-kill, Norton-Simon, and $E_{max}$. In all cases, the sequencing of chemotherapy followed by surgery led to better outcomes than surgery followed by chemotherapy. However, it is key to note that in order to make a conclusive recommendation, all reasonable possibilities and assumptions need to be examined. When data is available, these assumptions can be learned from data, as we endeavour to show by adapting the UPINN method to this problem.

\section{Methods}\label{sec:methods}

We build on the methods in~\cite{podina2022pinn}. In this work, we apply UPINNs in three different scenarios. First, we apply it to learn the drug action of a cancer growth ODE, modifying the method for better performance. Using the method, we are able to obtain an approximation of the drug action in three ODEs with three different drug actions. Secondly, we apply the method to a situation where the parameters vary with dose. We learn the parameters as a function of the dosage, and interpolate between observed dosages. Finally, we learn the net proliferation rate of doxorubicin from in-vitro data, after validating the method on similar synthetic data.

In the following subsections, we first describe how to apply the method generally, and in the subsequent sections, we detail the specific applications.

\subsection{General approach}

Suppose an ODE in the following form, with $m$ variables, where $u: t \rightarrow \mathds{R}^m$ and $F,G: u,t \rightarrow \mathds{R}^m$:

\begin{equation}\label{eq:1}
    \frac{du}{dt} = F(u,t) + G(u,t)
\end{equation}

Suppose furthermore that $F$ can be evaluated for every possible input, but $G$ is unknown and so cannot be. Then given a dataset of $n$ points $\{t_i,u_i\}$ which satisfy the differential equation, we can use the UPINN method to approximate $G$ at these points, and at points that are linear combinations of pairs $(t_j,u_j)$ in the dataset. We do this by representing $G(u,t)$ with a neural network $G_{NN}$, taking $u$ and $t$ as input and returning a vector-valued output for each such combination. However, some of the outputs of $G$ may be a priori known to be zero, and as such don't need to be modeled by the neural network. A case of this can be seen in the following sections. Furthermore, although in~\eqref{eq:1}, $F$ and $G$ are assumed to be added, this assumption need not hold in order for the method to learn $G$. Instead, there may be a different (known) function combining $F$ and $G$ on the right-hand side of~\eqref{eq:1} (e.g. $F \cdot G$) and if $G$ is identifiable given the data, the UPINN method should still be able to learn $G$. However, more complex expressions tend to be more difficult for the method to learn.

As per the standard PINN approach, along with an approximation of $G$, we obtain an approximation of the differential equation solution $u$ using a neural network $U_{NN}$. $U_{NN}$ and $G_{NN}$ are trained by minimizing the following loss:

\[\mathcal{L} = \mathcal{L}_{MSE} + \mathcal{L}_{ODE}\]

The MSE loss ensures that the output of the surrogate solution $U_{NN}$ adheres to the data $\{t_i,u_i\}$:

\[\mathcal{L}_{MSE} = \frac{1}{N}\sum_{i=1}^n (U_{NN}(t) - u_i)\]

As for the ODE loss, $U_{NN}$ is autodifferentiated\footnote{A common operation in the training of neural networks, which in this case computes the derivative of the output of the neural network.} with respect to $t$ and evaluated at a set of collocation points $\{t_j\}$. This loss minimizes the difference between the derivative of $U_{NN}(t)$ and the right hand side of eq.~\ref{eq:1}. In effect, this loss component enforces the differential equation to hold.

\[\mathcal{L}_{ODE} = \frac{1}{K}\sum_{j=1}^K \Bigg(\left.\frac{dU_{NN} (t)}{dt}\right|_{t_j} - (F(U_{NN}(t_j),t_j) + G_{NN}(U_{NN}(t_j),t_j))\Bigg)\]

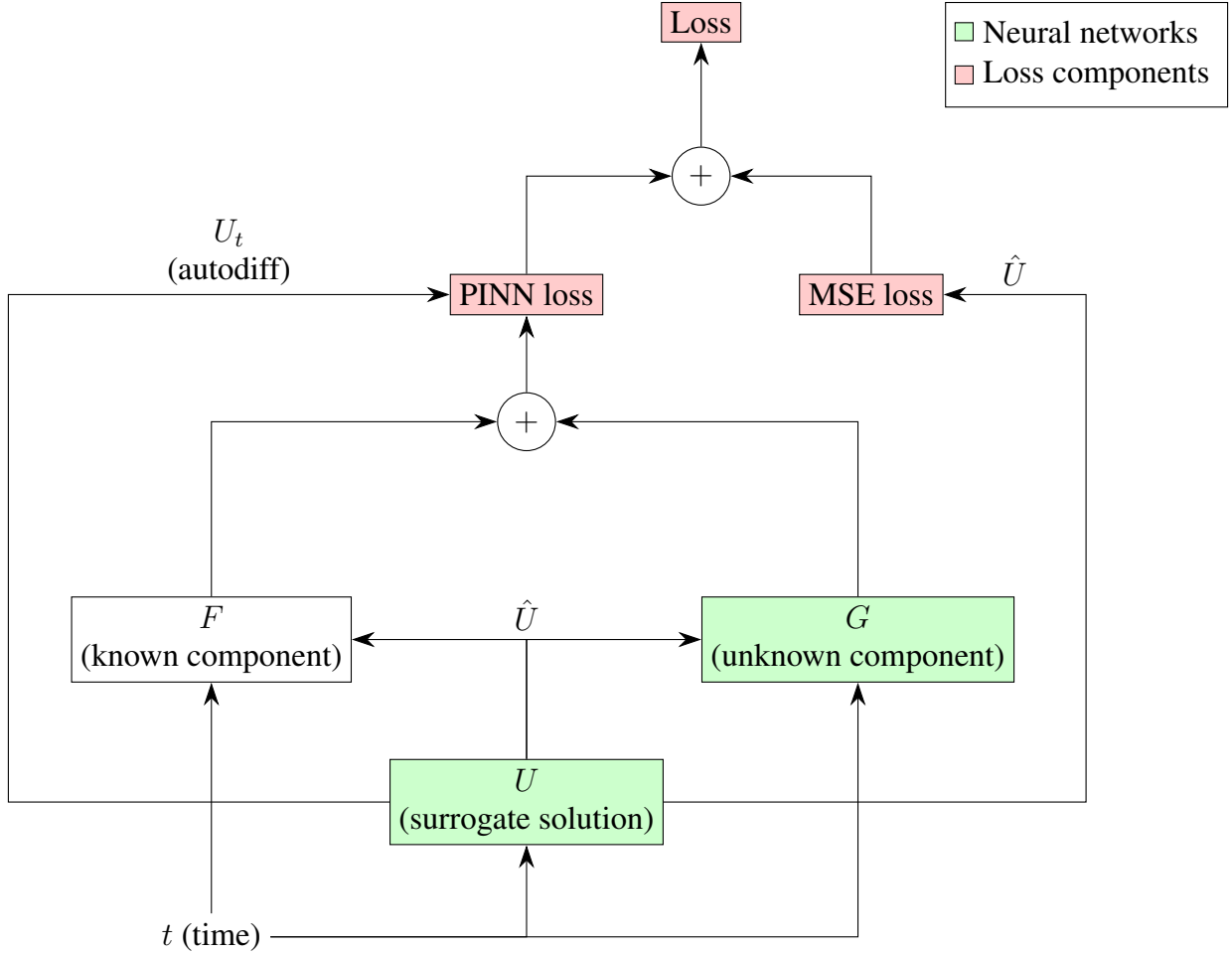
\begin{figure}[ht]
    \centering
\resizebox{\columnwidth}{!}{
  \begin{tikzpicture}[greennode/.style={shape=rectangle, fill=green, draw=black, fill opacity=0.2,text opacity=1.0,align=center},
  rednode/.style={shape=rectangle, fill=red, draw=black, fill opacity=0.2,text opacity=1.0,align=center}
  ,font=\sffamily]
  \tikzstyle{every node}=[font=\large]
    \node[circle,draw] (add_1) at (0,0) {$+$};
    \node[rednode]  (loss) at (0,2)  {Loss};
    
    \node[rednode] (pinn_loss) [below left=1cm and 1cm of add_1]  {PINN loss};% 2cm below, 1cm to the left (optional)
    \node[circle,draw] (add_2) [below=1cm of pinn_loss] {$+$};
    \node[rednode] (mse_loss) [below right=1cm and 1cm of add_1] {MSE loss};
    \node[rectangle,draw,align=center]  (G) [below left=2cm and 2cm of add_2]  {$F$\\(known component)};
    \node[greennode]  (F) [below right=2cm and 2cm of add_2]  {$G$ \\(unknown component)};
    \node[rectangle,draw,align=center,fill=green, fill opacity=0.2,text opacity=1.0]  (U) [below=4cm of add_2]  {$U$ \\(surrogate solution)};
    \node  (t) [below=3cm of G]  {$t$ (time)};
    \draw [-{Stealth[length=3mm, width=2mm]}] (mse_loss) |- (add_1);
    \draw [-{Stealth[length=3mm, width=2mm]}] (pinn_loss) |- (add_1);
    \draw [-{Stealth[length=3mm, width=2mm]}] (G) |- (add_2);
    \draw [-{Stealth[length=3mm, width=2mm]}] (F) |- (add_2);
    \draw [-{Stealth[length=3mm, width=2mm]}] (add_1) -- (loss);
    \draw [-{Stealth[length=3mm, width=2mm]}] (add_2) -- (pinn_loss);
    \draw [-{Stealth[length=3mm, width=2mm]}] (U.north) |- (F.west) node[midway,above] {$\hat{U}$} ;
    \draw [-{Stealth[length=3mm, width=2mm]}] (U.north) |- (G.east) ;
    \draw [-{Stealth[length=3mm, width=2mm]}] (t) -- (G);
    \draw [-{Stealth[length=3mm, width=2mm]}] (t.east) -| (U);
    \draw [-{Stealth[length=3mm, width=2mm]}] (t.east) -| (F);
    \draw [-{Stealth[length=3mm, width=2mm]}] (U.east) -| (5,-5) |- (mse_loss.east) node[pos=0.75,above,align=center] {$\hat{U}$};
    \draw [-{Stealth[length=3mm, width=2mm]}] (U.west) -| (-9,-5)  |- (pinn_loss.west) node[pos=0.75,above,align=center] {$U_t$ \\ (autodiff)};
    \matrix [draw, below] at (current bounding box.north east) {
      \node [greennode,label=right:Neural networks] {}; \\
      \node [rednode,label=right:Loss components] {}; \\
    };
    % \draw [-{Stealth[length=3mm, width=2mm]}] (U) -- (Y) node [midway,below,sloped] {*};
    % \draw [-{Stealth[length=3mm, width=2mm]}] (U) -- (Z) node [midway,below,sloped] {*};
    \end{tikzpicture}
    }
    \caption{Overview of the structure of the UPINN method as applied to~\eqref{eq:1}, which shows inputs and outputs of all known and unknown components, as well as losses. The surrogate solution $U$ outputted by the UPINN takes time $t$ as input. Both $F$ (the known component of the differential equation) and $G$ (the unknown component, to be fit by the UPINN) take in time and $\hat{U}$, the prediction of the neural network, as input.  $F$ and $G$, along with $U_t$ (the autodifferentiated derivative of $U_{NN}$ w.r.t. time) and is passed as input to the PINN loss. Then, the PINN loss computes the error between $U_t$ and $F+G$. The MSE loss computes the error between the surrogate solution $\hat{U}$ and the data.}
    \label{fig:NN_loss_diagram}
\end{figure}

$G_{NN}$ is trained through the ODE loss component, and $U_{NN}$ is trained through both loss components. A flowchart of the components and their interactions can be seen in Fig~\ref{fig:NN_loss_diagram}.

% talk about the two losses and how the training works

\subsection{Application to the identification of a cell-kill strategy}

We are interested in applying the above method to identify the kind of drug action that a specific chemotherapeutic has. Using an ODE, we simulate a situation where the chemotherapeutic is applied in a single dose to a culture of cells, and the cell numbers are measured at intervals after the application of the chemotherapeutic. We choose the following differential equation to simulate the concentration and cell numbers over time:
\begin{align}\label{eq:cell_kill_3_types}
    \frac{dN}{dt} &= \beta N (1-N) - C(t) G(N) \\
    \frac{dC}{dt} &= -\gamma N C
\end{align}

where $C(t)$ is the concentration of the drug (between 0 and 1), and $N(t)$ is normalized cell count, and as such takes values between 0 and 1. The first term governing the tumour dynamics is a growth term and is known. In general, we assume to have no information about $G(N)$, the drug action of the chemotherapeutic. However, for the purposes of testing the method, we choose $G$ to have one of the three following forms, as per~\cite{kohandel2006mathematical, panetta2003optimal}:

% is generated for a specified time interval:

% where $G(N)$, the drug action, is one of:

\begin{enumerate}
    \item $a_1 N$ (Log-kill model)
    \item $a_2 \beta N (1 - N)$ (Norton-Simon model)
    \item $a_3 N / (N + \delta)$ ($E_{max}$ model)
\end{enumerate}

where $a_j$ are constants. 

We generate synthetic data $\{t_i,N_i,c_i\}$ satisfying the above differential equation, with initial condition $N(0)=0.01$. $C$ is assumed to be zero for $t \in [0,12]$, and at day 12, the drug is instantaneously added to create a concentration of 1. From there the drug decays depending on the parameter $\gamma = 1.0$. For the $E_{max}$ case, we have $\delta=0.55$. $\beta$ is assumed to be known and equal to 1.0, but can be fit using a regular PINN from untreated growth data. Finally, the constants $a_{1,2,3}$ are set as $a_1 = 2.8, a_2=11.0, a_3=2.4$. An example of this workflow is shown in the results.

We set up three neural networks as follows: let $G(t,N) = C(t)G(N)$ and we approximate $G(t,N)$ with a neural network $G_{NN}$ with two inputs ($t$ and $N$) and one output. Separately, we approximate $C(t)$ by $C_{NN}$ with one input, $t$, and one output, $C(t)$. Finally, we have a neural network $U_{NN}$ with one input ($t$) and one output ($N$), which approximates the solution $N(t)$.

The workflow proceeds in three steps. First, we train $C_{NN}$ using only an MSE loss with respect to the $\{t_i,c_i\}$ data. The purpose of this network is to enable the interpolation of the concentration between observed timesteps. Then, we train $G_{NN}(t,N)$ using the data  $\{t_i,N_i\}$ using exactly the method described in the previous subsection. Finally, having obtained predictions for a set of times $\{t_j\}$ from both $C_{NN}$ and $G_{NN}$ we divide the prediction of $G_{NN}$ by the prediction of $C_{NN}$ to obtain an approximation of $G(N)$ for every timestep $t_j$. Since we also have the solution $U_{NN}$ for every timestep, we are able to get an approximation of $G(N)$ for any $N$ within the bounds of the data. This function cannot extrapolate to unseen $N$, so in generating 
$\{t_i,N_i,c_i\}$ it is important that the widest range possible is covered by the values $N_i$.

This method enables us to create a black-box representation of drug actions. The approximation of $G(N)$ for different timesteps $t_j$ may then be run through a symbolic regression algorithm to obtain a closed-form of the function. For example, AI Feynman~\cite{udrescu2020ai} can suggest possible closed-form functions linking the $t_j$ and $G(N)$, trading off the simplicity of the function with how well it fits the data. This is different than training a regression model or neural network because the symbolic regression algorithm searches a wide space of possible functions (e.g. including mathematical functions such as $\log$ or $\cos$ directly) rather than restricting the search space to polynomials or multi-layer perceptrons. Although the symbolic regression algorithm can find a closed form, which may shed some light on the underlying biological mechanisms of the hidden term, certain downstream applications for the drug action (e.g., finding an optimal drug schedule using reinforcement learning as in~\cite{eastman2021reinforcement}) do not require a closed form.

 The UPINN was implemented in PyTorch~\cite{paszke2019pytorch}. The neural network which represents the surrogate solution was a fully-connected network with 8 hidden layers with 20 hidden units each, tanh activation. All neural networks used in this work are structured similarly and use tanh activation. Each neural network which represents a hidden term was structured the same way. First, 5000 Adam~\cite{kingma2014adam} iterations were run, and then the Limited Memory BFGS optimization algorithm~\cite{liu1989limited} was run until sufficient convergence. The MSE and PINN loss components were weighted equally at all times. The inputs to each neural network were scaled such that it would be between 0 and 1 for all inputs (both time and number of cells).

\subsection{Application to parameter identification for several drug dosages}

In some situations, the differential equations that govern a specific biological process has parameters that vary not in time, but as a function of chemotherapy dosage. One such example can be created by assuming that the parameters $k_p, \theta$ in logistic growth

\begin{align}\label{eq:normal_logistic}
    \frac{dN}{dt} = k_p N \left(1 - \frac{N}{\theta}\right)
\end{align}

are in fact a function of the administered chemotherapy dosage, $D$. This changes the previous equation to

\begin{align}\label{eq:partial_logistic}
    \frac{\partial N(t,D)}{\partial t} = k_p(D) N(t,D) \left(1 - \frac{N(t,D)}{\theta(D)}\right)
\end{align}

where $k_p(D)$ and $\theta(D)$ are now the dose-dependent growth rate and carrying capacity. An initial condition constraint $N(0,D) = N_0$ can be included as well. In this case, for a higher dosage of chemotherapy, the growth of the cells would remain logistic, but the growth rate and carrying capacity would decrease. Conversely, if the dosage is low then the growth rate and carrying capacity would be higher. We consider this a realistic scenario for modelling the effects of doxorubicin, due to a model employed in a recent work~\cite{mckenna2017predictive}. 

Rather than solving~\eqref{eq:partial_logistic}, it is simpler to use an ODE solver to solve~\eqref{eq:normal_logistic} for different $(k_p,\theta)$. This generates several time series datasets $\{t_i,n_i\}_D$ of the number of tumour cells $n_i$ over time ($t$), one for each chemotherapy dosage condition $D$. It is possible to aggregate them into one dataset and fit $k_p(D)$ and $\theta(D)$ simultaneously to all available data. In our UPINNs setup, we simulate three different datasets using logistic growth, under low, medium and high dosage conditions. Three different dosages were considered in each experiment (low, medium and high), given as 15.0, 25.0, and 45.0 nM respectively to the model. Since this data is synthetic and the dosage is not used in the ODE directly, the numerical values of the drug concentration are somewhat arbitrary. 48 datapoints were used in total for each experiment, which means there were 16 datapoints per dosage. 
We added noise proportionally to the mean of the data, as per synthetic experiments in~\cite{podina2022pinn} at a noise level of 0.03. We aggregate the data for all dosages, and model the system via eq. \ref{eq:partial_logistic}. In our setup, we have three neural networks: one that models the surrogate solution, and two that model $k_p(D)$ and $\theta(D)$ respectively. The surrogate solution takes both the time and dosage as input. The remaining two neural networks have only one input (dosage) and one output each. As there are only three dosages for which $k_p$ and $\theta$ need to be fit, the parameters for unobserved dosages can be interpolated by the model. We note that additional constraints on the interpolation can be included by adding another loss term.

It is worth noting that if standard PINNs were utilized to fit these dose-dependent parameters, a separate PINN would have to be fit for the dataset corresponding to each dose. This is a more significant computational burden, and also not scalable when more than a few dosages are available. It also does not allow interpolation between dosages. By modelling eq (2) using UPINNs, we are able to fit one surrogate solution $N$ to all the datasets and obtain all the dose-dependent parameters simultaneously, provided they are identifiable.

The architecture of all neural networks was the same as in the previous section. Training proceeded very similarly, with 5000 Adam iterations at first. Then, the data loss was weighted with a value $\lambda=0.001$ and Adam training was continued for 1000 iterations. Then, L-BFGS finished the optimization. This training method was used in order to ensure that the PINN loss was minimized sufficiently during training. Similar scaling was employed on the input.

% Write about the PyTorch implementation etc.

\subsection{Application to in-vitro experimental data}

We now apply our method to identify the drug action of a real chemotherapeutic from in-vitro cell counts. For this purpose, we gathered the data used in~\cite{mckenna2017predictive}. In this series of experiments, four different cell lines (MDA-MB-468 (basal-like 1), SUM-149PT (basal-like 2), MDA-MB-231 (mesenchymal), and MDA-MB-453) were first allowed to grow undisturbed for at least three days, and then they were exposed to different concentrations of doxorubicin. The cells were exposed to doxorubicin for 6, 12 or 24 hours, after which the medium was changed. The concentrations of doxorubicin that were used by~\cite{mckenna2017predictive} are: 10nM, 20nM, 39nM, 78nM, 156nM, 312nM, 625nM, 1250nM, 2500nM.

In this work, the time-series cell count data is described by the following model:

\begin{align}
\frac{dN_{TC}}{dt} &= (k_p - k_d(t,D))N_{TC}(t)\left(1 - \frac{N_{TC}}{\theta(D)}\right)\label{eq:4}\\ % dN/dt with eqs 5A and 5B
k_d(t,D) &= k_{d,A}(D)\label{eq:5A} \\
k_d(t,D) &= k_{d,B}(D)r(D)te^{1-r(D)t}\label{eq:5B}
\end{align}

where $N_{TC}(t)$ is the number of cells over time, $k_p$ is the constant growth rate under control conditions, $k_d(D,t)$ is the death rate, dependent on dosage and time, $\theta(D)$ is the dose-dependent carrying capacity, and $r(D)$ is also a function of dosage. These parameters also take on different values per cell line. In~\cite{mckenna2017predictive}, eq~\ref{eq:5A} and eq~\ref{eq:5B} are fit separately, and the final prediction for each dosage is a weighted combination of the predictions of the two models. The authors have fit $k_{d,A}(D),r(D),k_{d,A}(D)$ as constant parameters per dosage using a nonlinear least squares approach. No time-dependent or dose-dependent functions are fit.

We note that if $k_p$ is allowed to take on any value, then fitting $k(D,t)$ and $\theta(D)$ simultaneously using our method is not identifiable. Hence, we simplified this model to the two following forms:
\begin{align}\label{eq:F}
    \frac{dN_{TC}(t)}{dt} = F_D(N,t) N_{TC}(t)%\tag{7} 
\end{align}
\begin{align}\label{eq:G}
    \frac{dN_{TC}(t)}{dt} = G_D(t) N_{TC}(t) \left( 1 - N_{TC}(t))\right)%\tag{8}
\end{align}

The subscript $D$ indicates that this function is different for each dosage. In our implementation, $F_D$ takes only time as the input, although it could take both numbers of cells, $N$ and time $t$ as input. We suggest and work with these two formulations of the hidden term, because for different applications, more prior knowledge can be incorporated if it is known.

Equations~\ref{eq:F} and~\ref{eq:G} are certainly identifiable for all tuples of time, cell count and time derivative values, denoted by $\{t_i,N_i,dN_i\}$. This is because for every tuple $\{t_i,N_i,dN_i\}$ at each of the collocation points, there is a unique solution of either eqs.~\ref{eq:F} and~\ref{eq:G} for $F_D$ or $G_D$ (respectively) at each dose $D$. Note that the equations~\ref{eq:F} and~\ref{eq:G} are general enough to be applicable to several scenarios: simulated data from eq~\ref{eq:5A}, simulated data from eq~\ref{eq:5B}, and the in-vitro data. Hence, we learn $F_D(N,t)$ and $G_D(t)$ for all three of these cases, except $F_D(N,t)$ generated using eq~\ref{eq:5A}. This is because $F_D(N,t)$ takes on a single constant value in that case.

The loss function was the same as in the previous sections, and the training proceeded similarly. The neural network architectures used were of 8 hidden layers of 128 units each, for each fully-connected neural network. Training was done such that Adam iterations were stopped at 1500 iterations, or upon reaching a loss of $2\cdot 10^{-3}$, whichever comes first. Then, L-BFGS iterations were started. This conditional stopping of Adam iterations was done in order to bring the weights sufficiently close to the optimum before L-BFGS starts, but not so close that L-BFGS fails to perform any optimization steps.

\section{Results}\label{sec:results}

In the following three sections, we detail the performance of the identification of hidden terms in all three of the scenarios. We apply our method to learn the drug action (the model that best describes the cell dynamics after treatment) as a function of the cell count only, then we learn growth rate parameters as a function of the dosage, and finally we learn the net proliferation rate of doxorubicin as a function of time. 

\subsection{Identification of drug action (Log-kill, NS, E-max) from synthetic data}

As described in the methods, we generate synthetic data for three different drug actions, and we evaluate how well the underlying true drug action can be recovered. We test our method on sparse noiseless data generated in two different ways, and noisy data. We also test a two-step procedure where we recover $\beta$ first from the untreated data using a regular PINN, and then apply our method to find the drug action.

% interesting: pick data points based on the difference between N... keep a minimum distance... generate lots first and then throw them away.

\begin{figure}[ht]
    \centering
    \subfigure[]{\includegraphics[width=0.3\textwidth]{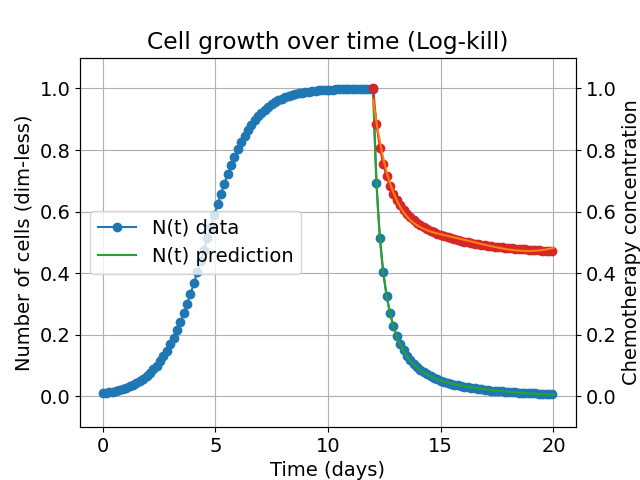}}
    \subfigure[]{\includegraphics[width=0.3\textwidth]{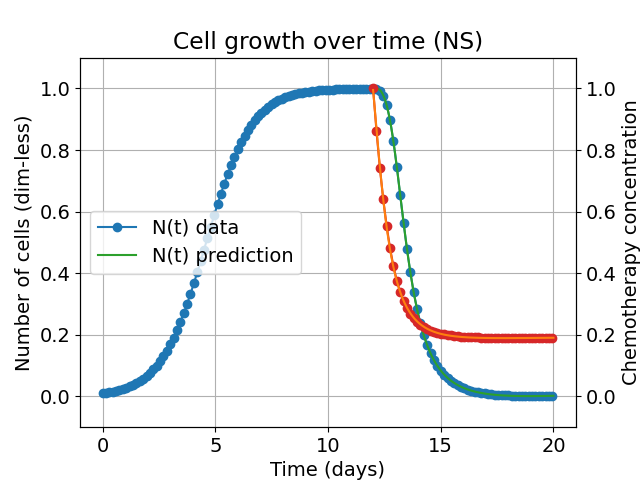}} 
    \subfigure[]{\includegraphics[width=0.3\textwidth]{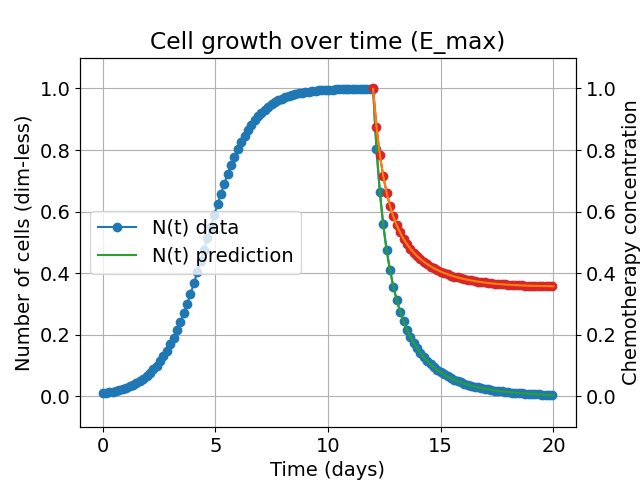}}
    \subfigure[]{\includegraphics[width=0.3\textwidth]{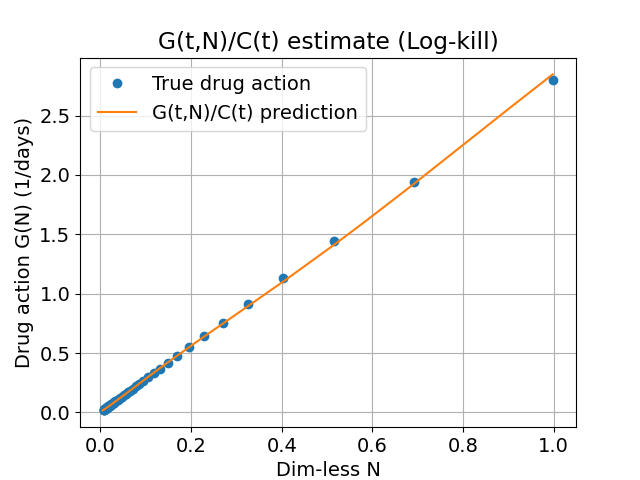}}
    \subfigure[]{\includegraphics[width=0.3\textwidth]{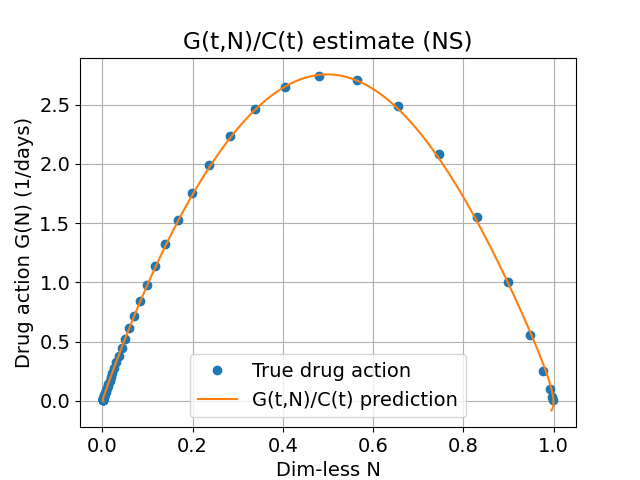}} 
    \subfigure[]{\includegraphics[width=0.3\textwidth]{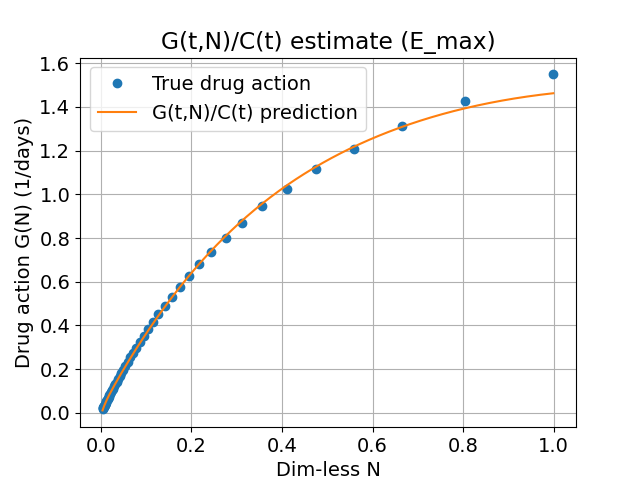}}
    \caption{Datasets for the different drug actions (a, b, c) with their respective drug actions learned below them (d, e, f). Data was collected with a fixed time period (0.15 days) elapsing between each data point. The drug actions are the functions $G(N)$ in Eq~\eqref{eq:cell_kill_3_types}, replaced by one of: log-kill, Norton-Simon and $E_{max}$.}
    \label{fig:equispace}
\end{figure}

First, we test on sparse data with spacing that is uniform through time. We generate a time series dataset of 54 points for each drug action. Fig~\ref{fig:equispace} shows the learned drug action for all three types of cell-kill strategies, when the spacing was uniform in time. In other words, it plots the learned $G(N)$ as a function of different $N$ that appeared in the training data. We note that the mean squared error (MSE) between the true drug action and the predicted drug action, as reported in Table~\ref{tab:section_1_mses} for all three types of drug action, are all on the order of $10^{-4}$. A visual inspection shows that the predicted drug action matches the true drug action for almost all values of $N$.
%Visually, the estimate matches the true drug action when $N$ is close to 0, as $N$ increases the estimate begins to slightly deviate from the data. This mainly appears to be due to the lack of data for values of $N$ close to 1. Examining the plots of the data, one can see that for high values of $N$ (which occur right after the administration of the chemotherapeutic), the rate of tumour cell death is rapid, which leads to less data being collected during that time. For this reason, we also constructed the dataset (with the same parameters) in a different way, where we adjusted the spacing of the data to account for high initial rates of cell decline. 

% \begin{figure}
%     \centering
%     \subfigure[]{\includegraphics[width=0.3\textwidth]{figures/data_num_11_eq_space.png}}
%     \subfigure[]{\includegraphics[width=0.3\textwidth]{figures/data_num_12_eq_space.png}} 
%     \subfigure[]{\includegraphics[width=0.3\textwidth]{figures/data_num_13_eq_space.png}}
%     \subfigure[]{\includegraphics[width=0.3\textwidth]{figures/n_fun_equispace_logkill_model2.png}}
%     \subfigure[]{\includegraphics[width=0.3\textwidth]{figures/n_fun_equispace_NS_model2.png}} 
%     \subfigure[]{\includegraphics[width=0.3\textwidth]{figures/n_fun_equispace_Emax_model2.png}}
%     \caption{Datasets for the different drug actions (a, b, c) with their respective drug actions learned below them (c, d, f). Data was collected with a fixed time period elapsing between each data point.}
%     \label{fig:equispace}
% \end{figure}

\begin{figure}[ht]
    \centering
    \subfigure[]{\includegraphics[width=0.3\textwidth]{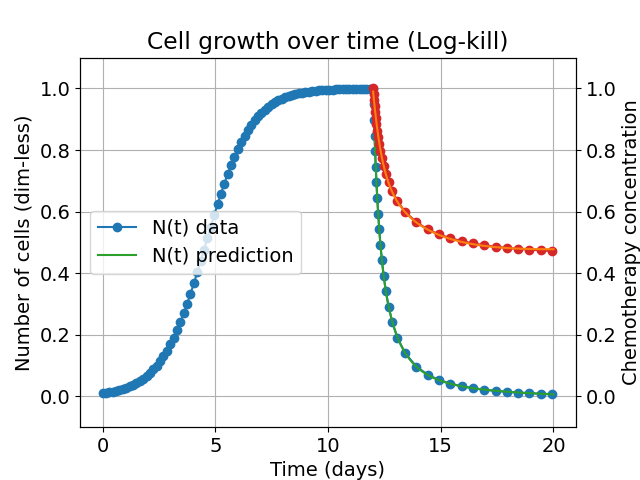}}
    \subfigure[]{\includegraphics[width=0.3\textwidth]{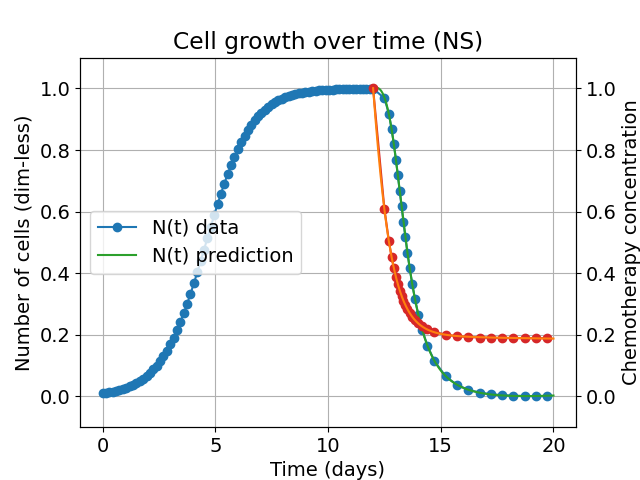}} 
    \subfigure[]{\includegraphics[width=0.3\textwidth]{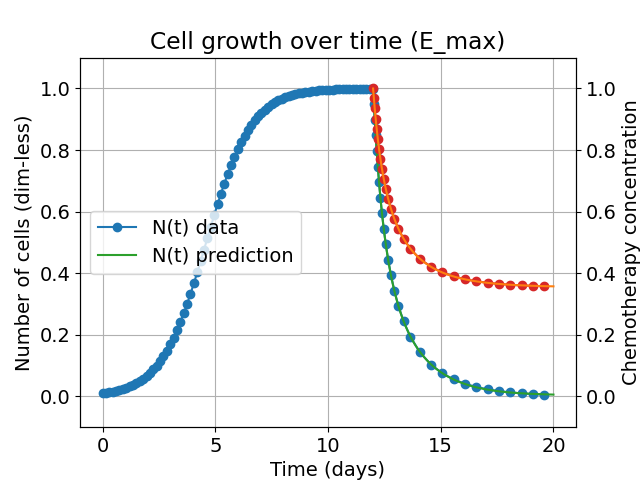}}
    \subfigure[]{\includegraphics[width=0.3\textwidth]{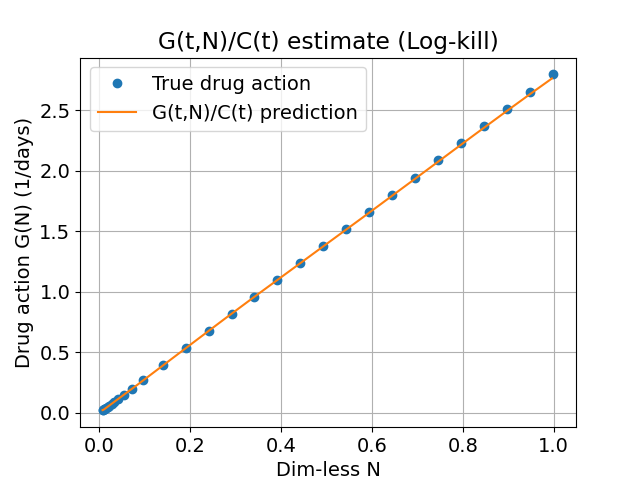}}
    \subfigure[]{\includegraphics[width=0.3\textwidth]{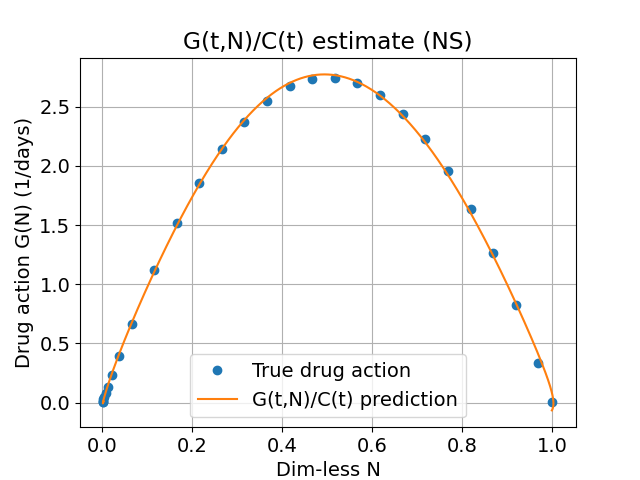}} 
    \subfigure[]{\includegraphics[width=0.3\textwidth]{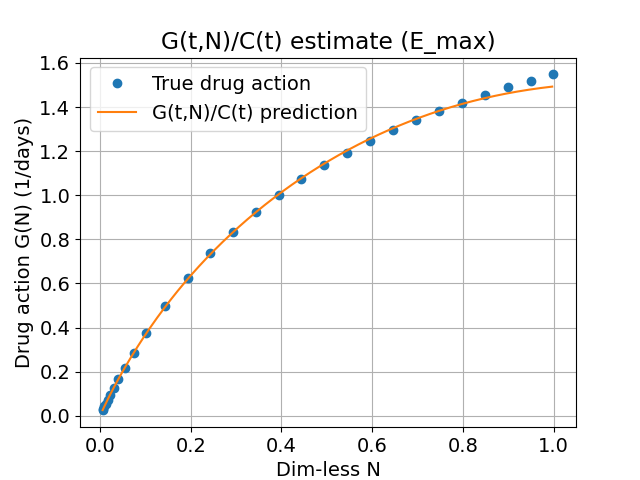}}
    \caption{Datasets for the different drug actions (a, b, c) with their respective drug actions learned below them (d, e, f). The data is noiseless but collected such that there is at one datapoint for each 0.05-interval decrease in $N$. Although there are half as many points collected than in the equispaced case, the MSE between the true drug action and the predicted drug action is still on the order of $10^{-4}$ as shown by Table~\ref{tab:section_1_mses}}
    \label{fig:adjusted}
\end{figure}

For the second data collection strategy, the spacing of the datapoints was dynamically chosen based on the rate of change of cell populations, so that there are proportionally more points at higher-slope areas compared to low-slope areas. This was done by first solving each ODE with many timesteps (time-step size 0.001), and then filtering the resulting data. The settings for the filtering were such that 1 point was present for each cell decline of 0.05, and no more than 500 contiguous points were removed. This yielded datasets of 31 points for the log-kill model, and 29 points for the others. The results of running the UPINNs on these datasets can be seen in Fig~\ref{fig:adjusted}. As evidenced by the mean-squared errors in Table~\ref{tab:section_1_mses}, the learned drug actions for both the log-kill and Norton-Simon model are on the order of $10^{-4}$ and $10^{-5}$, giving comparable or better MSEs than using spacing that is equal in time. Given that the adjusted spacing uses half as much data, but achieves errors on the same order of magnitude ($10^{-4}$) highlights the importance of data collection in high-slope areas. In practice, this might mean collecting data as frequently as possible when a high cell decline is to be expected, and collecting less frequently when the cell growth has reached a plateau.

% \begin{figure}
%     \centering
%     \subfigure[]{\includegraphics[width=0.3\textwidth]{figures/data_num_4_special_space.png}}
%     \subfigure[]{\includegraphics[width=0.3\textwidth]{figures/data_num_5_special_space.png}} 
%     \subfigure[]{\includegraphics[width=0.3\textwidth]{figures/data_num_6_special_space.png}}
%     \subfigure[]{\includegraphics[width=0.3\textwidth]{figures/n_fun_logkill_model2.png}}
%     \subfigure[]{\includegraphics[width=0.3\textwidth]{figures/n_fun_NS_model2.png}} 
%     \subfigure[]{\includegraphics[width=0.3\textwidth]{figures/n_fun_Emax_model2.png}}
%     \caption{Datasets for the different drug actions (a, b, c) with their respective drug actions learned below them (c, d, f). Data collection was adjusted to be more frequent in areas of higher cell decline.}
%     \label{fig:adjusted}
% \end{figure}

% \begin{figure}
%     \centering
%     \subfigure[    \label{fig:foobar_1}]{\includegraphics[width=0.48\textwidth]{figures/N_type_1.png}} 

%     \subfigure[]{\includegraphics[width=0.48\textwidth]{figures/N_type_2.png}} 
%     \subfigure[]{\includegraphics[width=0.48\textwidth]{figures/N_type_3.png}}
%     \caption{(a) 31 points (b) 28 points (c) 29 points}
%     \label{fig:foobar}
% \end{figure}

% \begin{figure}
%     \centering
%     \subfigure[]{\includegraphics[width=0.48\textwidth]{figures/N_type_1.png}}
%     \subfigure[]{\includegraphics[width=0.48\textwidth]{figures/N_type_2.png}} 
%     \subfigure[]{\includegraphics[width=0.48\textwidth]{figures/N_type_3.png}}
%     \caption{(a) 31 points (b) 28 points (c) 29 points}
%     \label{fig:foobar}
% \end{figure}

\begin{figure}[ht]
    \centering
    \subfigure[]{\includegraphics[width=0.3\textwidth]{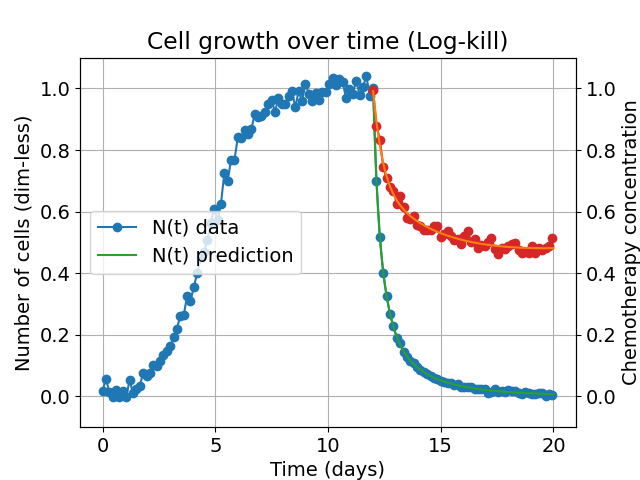}}
    \subfigure[]{\includegraphics[width=0.3\textwidth]{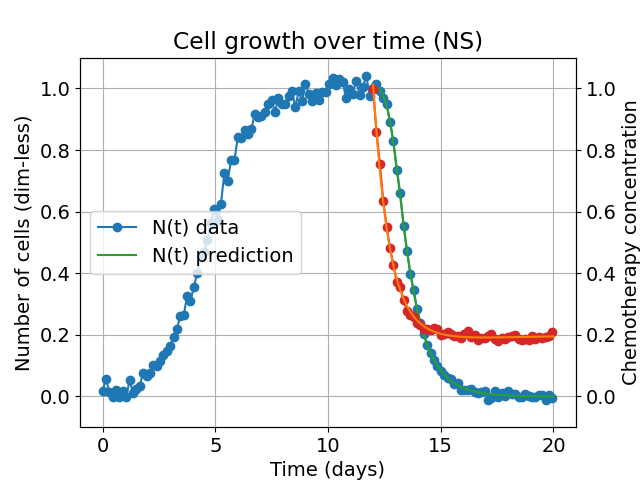}} 
    \subfigure[]{\includegraphics[width=0.3\textwidth]{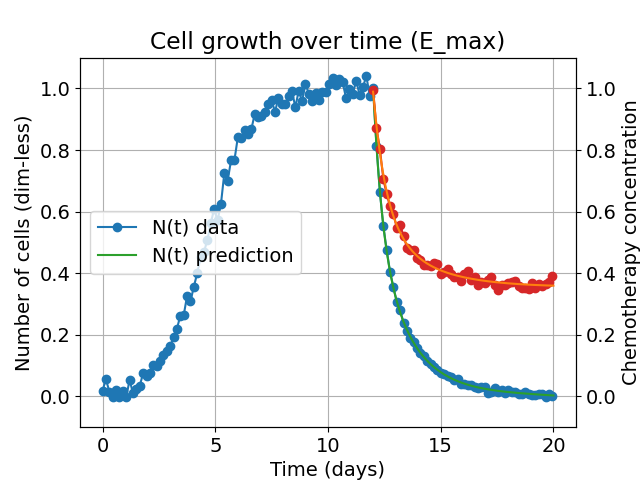}}
    \subfigure[]{\includegraphics[width=0.3\textwidth]{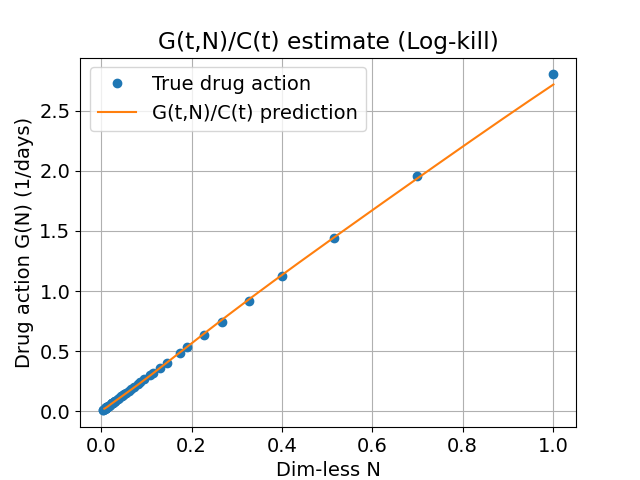}}
    \subfigure[]{\includegraphics[width=0.3\textwidth]{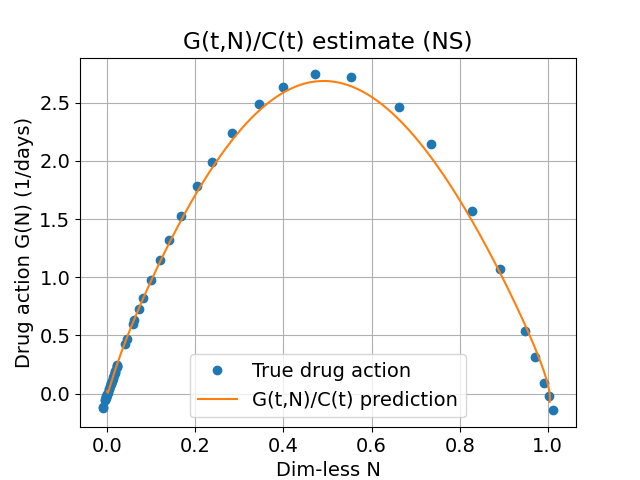}} 
    \subfigure[]{\includegraphics[width=0.3\textwidth]{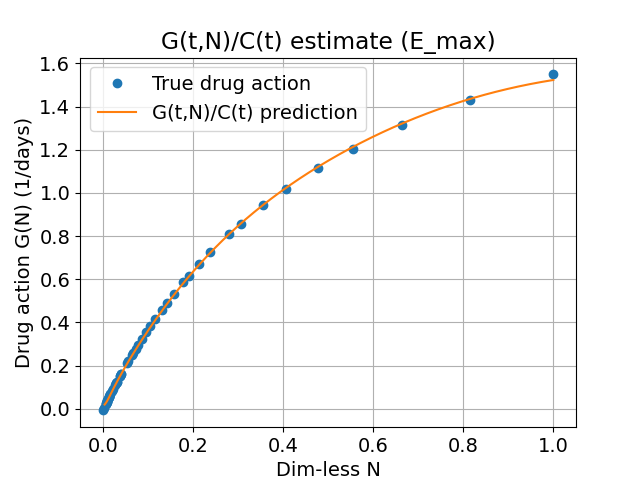}}
    \caption{Datasets for the different drug actions (a, b, c) with their respective drug actions learned below them (c, d, f). The data has a noise level of 0.03 added proportionally to the mean of the variables.}
    \label{fig:adjusted_noisy}
\end{figure}

\begin{figure}[ht]
    \centering
    \subfigure[]{\includegraphics[width=0.3\textwidth]{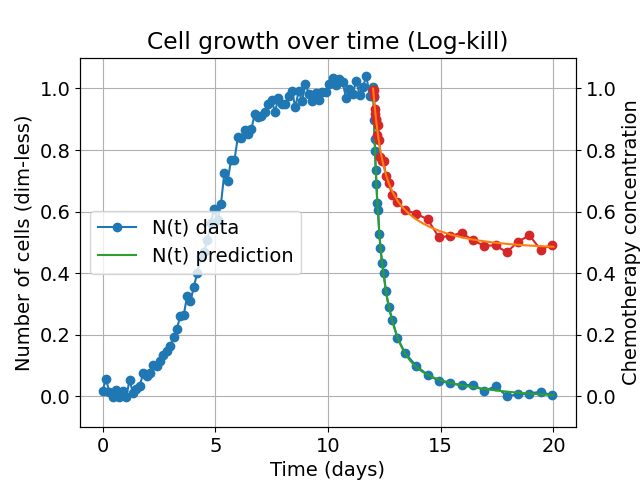}}
    \subfigure[]{\includegraphics[width=0.3\textwidth]{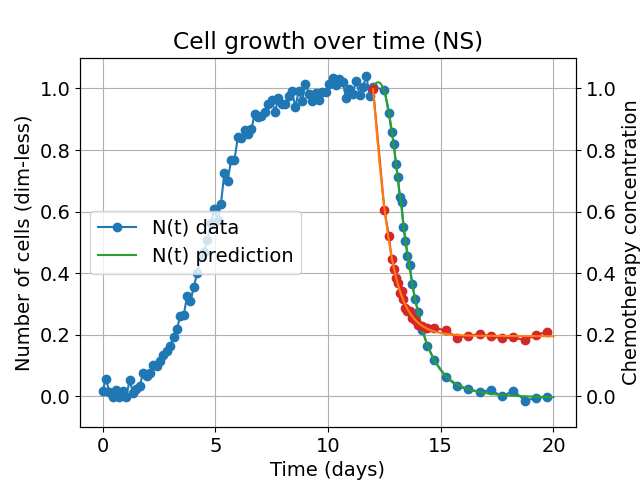}} 
    \subfigure[]{\includegraphics[width=0.3\textwidth]{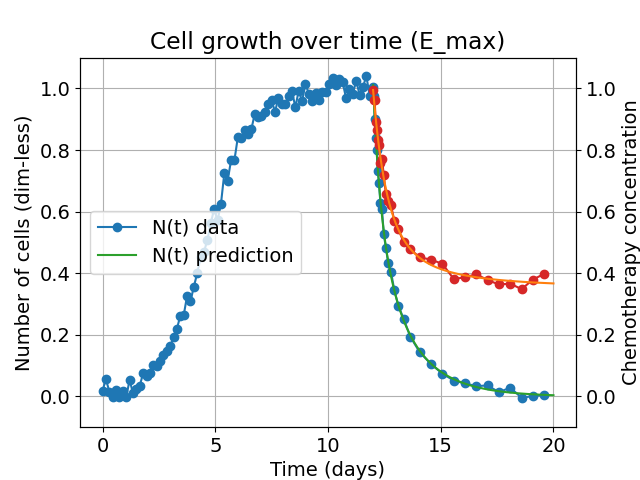}}
    \subfigure[]{\includegraphics[width=0.3\textwidth]{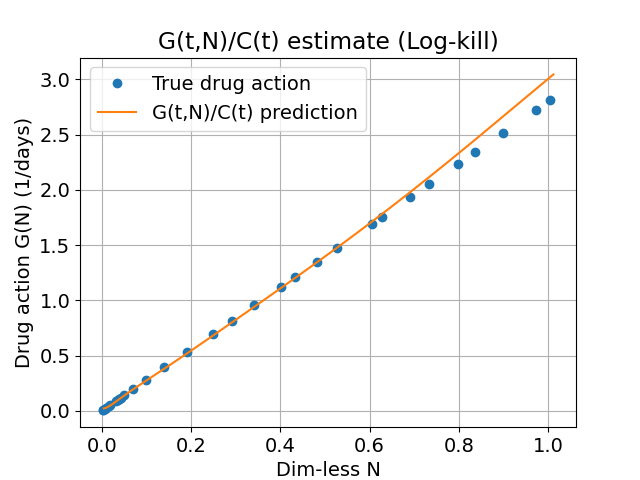}}
    \subfigure[]{\includegraphics[width=0.3\textwidth]{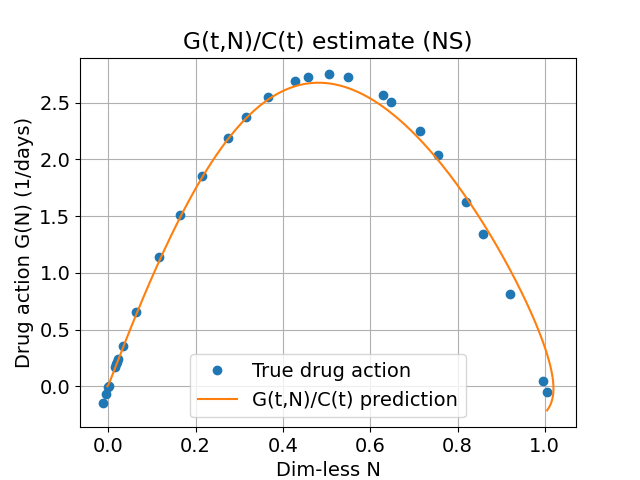}} 
    \subfigure[]{\includegraphics[width=0.3\textwidth]{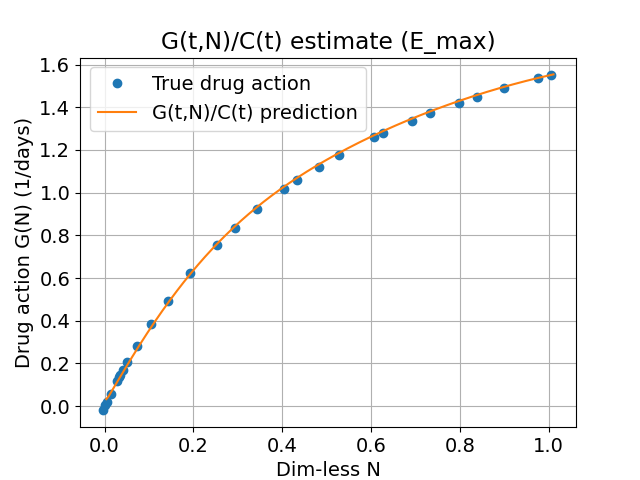}}
    \caption{Datasets for the different drug actions (a, b, c) with their respective drug actions learned below them (c, d, f). The data is spaced so that times of high cell decline have proportionally more observations (one datapoint for each 0.05-interval decrease in $N$). It also has a noise level of 0.03 added proportionally to the mean of the variables.}
    \label{fig:equispaced_noisy}
\end{figure}

Then, we test the method on noisy data, constructed by adding noise to both dynamically spaced (Fig~\ref{fig:adjusted_noisy}) and equally spaced data (Fig~\ref{fig:equispaced_noisy}). For each of the drug actions (log-kill, NS, and $E_{max}$), we find that the method has an MSE of $10^{-4}$, $10^{-3}$, and $10^{-5}$ respectively (Table~\ref{tab:section_1_mses}) under noisy conditions when the spacing is equal in time, but not when the spacing is adjusted dynamically. When the spacing is adjusted dynamically then the errors are an order of magnitude higher ($10^{-3}$, $10^{-2}$, and $10^{-4}$ respectively). The noise is added proportionally to the mean of the data, with the noise level being $0.03$. The predicted drug action is visually quite a bit different from the true $G(N)$ for larger values of $N$ when the spacing is adjusted. More investigation is needed into why the adjusted spacing has a detrimental effect when the data is noisy. In addition, it is worth noting that adding only noise to equally spaced data has a negligible effect on the MSE of all drug actions. The MSEs remain at most $10^{-4}$ for both noisy and noiseless equispaced data. This highlights the robustness of the method. A comparison of the MSE values of the learned drug actions for all three types of tests can be seen in Table~\ref{tab:section_1_mses}, with the lowest values for each column bolded.

% \begin{figure}
%     \centering
%     \subfigure[]{\includegraphics[width=0.3\textwidth]{figures/data_num_14_eq_space.png}}
%     \subfigure[\label{fig:noisy_2}]{\includegraphics[width=0.3\textwidth]{figures/data_num_17_eq_space.png}} 
%     \subfigure[]{\includegraphics[width=0.3\textwidth]{figures/data_num_16_eq_space.png}}
%     \subfigure[]{\includegraphics[width=0.3\textwidth]{figures/n_fun_noisy_logkill_model2.png}}
%     \subfigure[]{\includegraphics[width=0.3\textwidth]{figures/n_fun_noisy_NS_model2.png}} 
%     \subfigure[]{\includegraphics[width=0.3\textwidth]{figures/n_fun_noisy_Emax_model2.png}}
%     \caption{Datasets for the different drug actions (a, b, c) with their respective drug actions learned below them (c, d, f). The data is slightly noisy.}
%     \label{fig:noisy}
% \end{figure}

\begin{table}[ht]
\centering
    \def\arraystretch{1.5}%
\begin{tabular}{|l|l|l|l|l}

\cline{1-4}
\diagbox[width=4cm]{Type of data\kern-1.0em}{\kern-1.0em Drug action}          & Log-kill & Norton-Simon & $E_{max}$ &  \\ \cline{1-4}
Equal spacing       &     $1.00\times 10^{-4}$     &  $4.13\times 10^{-4}$  &     $1.85\times 10^{-4}$       &  \\ \cline{1-4}
Adjusted spacing    &      $\mathbf{6.30\times 10^{-5}}$    &  $\mathbf{3.84\times 10^{-4}}$  &    $2.00\times 10^{-4}$       &  \\ \cline{1-4}
Equal spacing + noise        &     $1.66\times 10^{-4}$     &   $4.17\times 10^{-3}$  &    $\mathbf{7.32\times 10^{-5}}$       & \\ \cline{1-4}
Adjusted spacing + noise        &     $4.29\times 10^{-3}$     &   $1.22\times 10^{-2}$  &    $1.96\times 10^{-4}$       & \\ \cline{1-4}

\end{tabular}
\vspace{1em}
\caption{Mean-squared error of the drug action predictions (compared to the ground truth drug action) for different types of data. Equal spacing means that the was equally spaced in time (one datapoint per 0.15 days). Adjusted spacing means that the data was collected with proportionally more datapoints when the cell decline has a high rate. When noise is added, the noise level is 0.03, added proportionally to the mean of the data. The lowest value in each column is bolded.}
\label{tab:section_1_mses}
\end{table}

% \begin{table}[]
% \centering
%     \def\arraystretch{1.5}%
% \begin{tabular}{|l|l|l|l|l}

% \cline{1-4}
% \diagbox[width=4cm]{Type of data\kern-1.0em}{\kern-1.0em Drug action}          & Log-kill & NS & $E_{max}$ &  \\ \cline{1-4} 
% Adjusted spacing    &      $4.04\times 10^{-3}$    &  $3.94\times 10^{-4}$  &    $1.73\times 10^{-4}$       &  \\ \cline{1-4}
% Equal spacing       &     $4.37\times 10^{-3}$     &  $6.86\times 10^{-4}$  &     $2.18\times 10^{-4}$       &  \\ \cline{1-4}
% Added noise         &     $5.28\times 10^{-3}$     &   $7.18\times 10^{-3}$  &    $1.38\times 10^{-3}$       & \\ \cline{1-4}

% \end{tabular}
% \vspace{1em}
% \caption{Mean-squared error of the drug action predictions for different types of data}
% \label{tab:my_label}
% \end{table}

\begin{figure}
    \centering
    \subfigure[]{\includegraphics[width=0.45\textwidth]{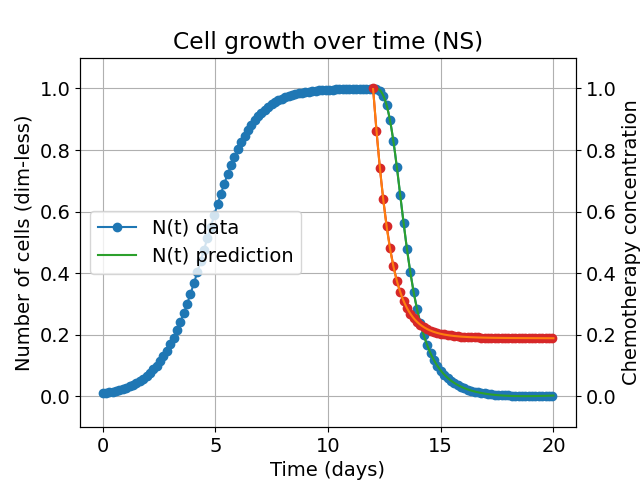}}
    \subfigure[]{\includegraphics[width=0.45\textwidth]{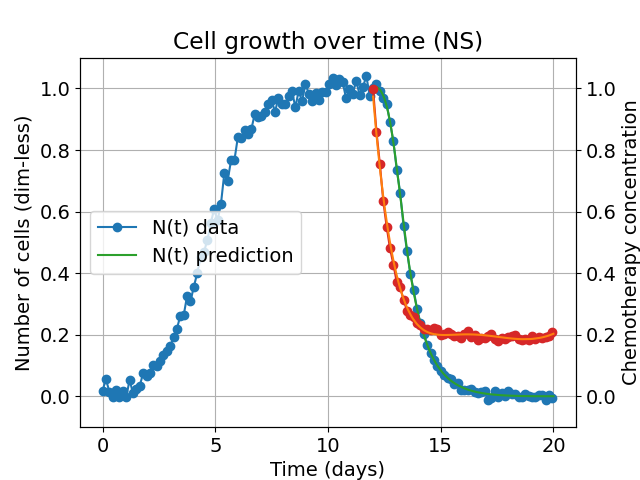}} 
    \subfigure[]{\includegraphics[width=0.45\textwidth]{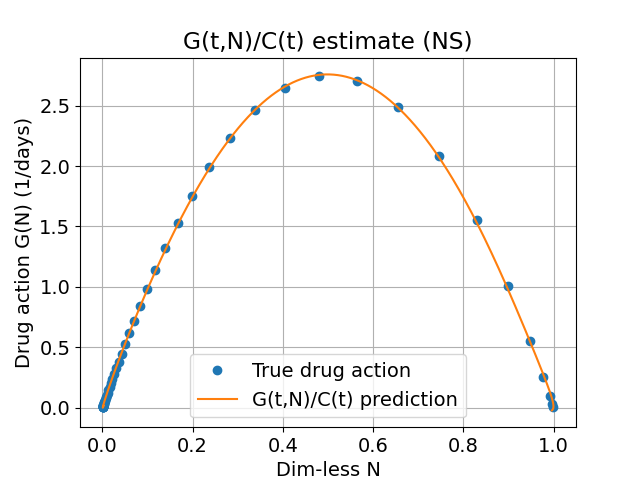}}
    \subfigure[]{\includegraphics[width=0.45\textwidth]{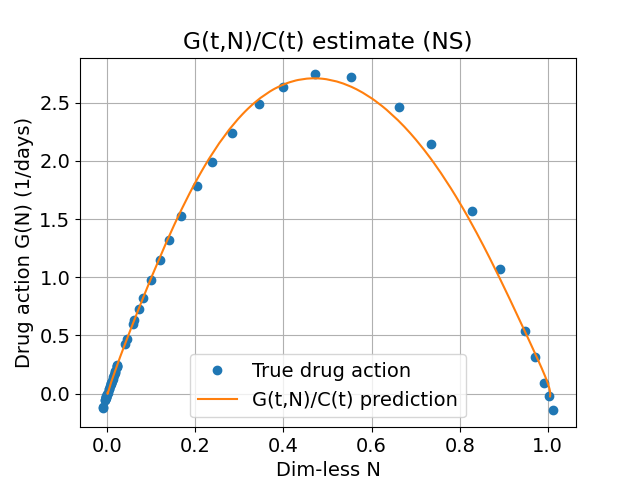}}
    \caption{ (a) Equispaced noiseless data with Norton-Simon drug action, (b) Learned drug action after the parameter $\beta$ was fit from the first 12 days of noiseless data (final estimate: 0.999, MSE $1\times 10^{-6}$),(c) Equispaced noisy (0.03) data with Norton-Simon drug action, (d) Learned drug action after the parameter $\beta$ was fit from the first 12 days of noisy data (final estimate: 0.997, MSE $9\times 10^{-6}$)}
    \label{fig:two_step}
\end{figure}

Finally, we apply a two-step process where first $\beta$ is learned using a standard PINN from both noiseless and noisy data, and then this $\beta$ estimate is employed during the estimation of $G(N)$. This showcases a way to apply UPINNs where no prior knowledge of the parameters or drug action is needed. The results of this test can be seen in Fig~\ref{fig:two_step}. Fig~\ref{fig:two_step} (a) is the noiseless data, created using the same parameters as before, using the Norton-Simon model. For noiseless data, we obtain an estimate of $\beta=0.999$ from the untreated data, which yields an MSE of $1\times 10^{-6}$ compared to the true $\beta=1$, with subsequent MSE of the drug action being $1.3\times 10^{-4}$. For noisy data, we obtain an estimate of $\beta = 0.997$ (MSE of $9\times 10^{-6}$), with subsequent MSE of the drug action being $4.5\times 10^{-3}$. The performance is clearly worse when the data is noisy, but a visual inspection shows that we can recover the curve and $\beta$ well for both noisy and noiseless data.

% \begin{figure}
%     \centering
%     \subfigure[]{\includegraphics[width=0.3\textwidth]{figures/data_num_31_two_step.png}}
%     \subfigure[]{\includegraphics[width=0.3\textwidth]{figures/two_step_Type_2_prediction.png}} 
%     \subfigure[]{\includegraphics[width=0.3\textwidth]{figures/n_fun_two_step_NS_model2.png}}
%     \caption{(a) Data generated as per Fig 4b (b) Surrogate solution learned on the interval of the untreated data (c) Learned drug action}
%     \label{fig:two_step}
% \end{figure}

Overall, we show that the UPINN method applied to the discovery of chemotherapeutic drug actions is effective in recovering the drug action in three common cases. It recovers the drug action with a low MSE ($10^{-3}$) in cases of both sparse and noisy data, and can be combined with the standard PINN approach to yield MSEs on the same order of magnitude.

\subsection{Inference and interpolation of dose-dependent logisitic growth parameters}

In this set of synthetic experiments, we show that we attain low MSE ($10^{-3}$ to $10^{-8}$) in recovering the true proliferation rate $k_p(D)$ and carrying capacity $\theta(D)$ for known dosages, and interpolate between the dosages. At the low dosage, the growth rate and carrying capacity are set to be the highest ($k_p=0.03,0.08$ $\text{hr}^{-1}$, $\theta=2.0,4.0$), and at the highest dosage they are the lowest ($k_p=0.01,0.06$ $\text{hr}^{-1}$, $\theta=1.0,3.0$). The set of values chosen for $k_p$ reflects either low growth rates (0.01-0.03), high growth rates overall (0.06-0.08), or a large range of growth rates (0.01-0.08). Similar parameter choices were made for $\theta$, where low growth rates correspond to $\theta$ ranges 1.0-2.0, large range of growth rates correspond to ranges 1.0-4.0, and high growth rates overall correspond to $\theta$ between 3.0-4.0. Table \ref{tab:noiseless_interp} and \ref{tab:noisy_interp} show the accuracy in recovering the true parameters and fitting the data for both noisy and noiseless data respectively. The true parameters are recovered on the order of $10^{-3}$ to $10^{-6}$ even in cases of sparse and noisy data. The specific values of these parameters were selected to be similar to those fit by~\cite{mckenna2017predictive}. The resulting surrogate solution MSE for noiseless and noisy data is at most $10^{-3}$.

\begin{table}[ht]
    \centering
    \def\arraystretch{1.5}%
    \begin{tabular}{|l|l||l|l|l|}
    \hline
         $k_p(D)$&  $\theta(D)$& MSE of $\theta$& MSE of $k_p$& Model MSE\\
         \hline
         $[0.03, 0.02, 0.01]$&  $[2.0, 1.5, 1.0]$&  0.00012& 6.3e-7&2.4e-05\\
         $[0.08, 0.07, 0.06]$&  $[2.0, 1.5, 1.0]$&  3.678e-06& 6.5e-05&0.0002\\
         $[0.08, 0.05, 0.01]$&  $[2.0, 1.5, 1.0]$&  0.00014& 0.0029& 0.00154\\
         $[0.03, 0.02, 0.01]$&  $[4.0, 3.5, 3.0]$&  0.000465&4.456e-08&2.8e-05\\
         $[0.08, 0.07, 0.06]$&  $[4.0, 3.5, 3.0]$&  4.94e-05& 6.69e-05&0.0004\\ 
         $[0.08, 0.05, 0.01]$&  $[4.0, 3.5, 3.0]$&   0.00042&1.67e-05&0.006\\
         $[0.03, 0.02, 0.01]$&  $[4.0, 2.5, 1.0]$&  0.0011&5.8e-07&1.9e-05\\
         $[0.08, 0.07, 0.06]$&  $[4.0, 2.5, 1.0]$&  4.35e-05&7.01e-05&0.0002\\
         $[0.08, 0.05, 0.01]$&  $[4.0, 2.5, 1.0]$&  0.0036&0.00089&0.0035\\
         \hline
    \end{tabular}
    \vspace{1em}
    \caption{Mean squared errors for each experiment (noiseless data). The row of $\theta$ and $k_p$ indicates the values used for each dosage (low, medium, and high dosages respectively). Model MSE is the MSE between the data and the model.}
    \label{tab:noiseless_interp}
\end{table}

\begin{table}[ht]
    \centering
    \def\arraystretch{1.5}%
    \begin{tabular}{|l|l||l|l|l|}
    \hline
         $k_p(D)$&  $\theta(D)$& MSE of $\theta$& MSE of $k_p$& Model MSE\\
         \hline
         $[0.03, 0.02, 0.01]$&  $[2.0, 1.5, 1.0]$&   0.0006&3.72e-06&0.001\\%
         $[0.08, 0.07, 0.06]$&  $[2.0, 1.5, 1.0]$&  0.0001&7.97e-05&0.0017\\%
         $[0.08, 0.05, 0.01]$&  $[2.0, 1.5, 1.0]$&  0.00024&0.0029&0.0015\\%
         $[0.03, 0.02, 0.01]$&  $[4.0, 3.5, 3.0]$& 0.0029&9.0e-07&0.0044\\%
         $[0.08, 0.07, 0.06]$&  $[4.0, 3.5, 3.0]$&  0.00016&6.6e-05&0.0086\\ %
         $[0.08, 0.05, 0.01]$&  $[4.0, 3.5, 3.0]$&   0.025&0.00094&0.0095\\%
         $[0.03, 0.02, 0.01]$&  $[4.0, 2.5, 1.0]$&   0.0011&2.7e-07&0.0033\\%
         $[0.08, 0.07, 0.06]$&  $[4.0, 2.5, 1.0]$&   0.00023&7.1e-05&0.0055\\%
         $[0.08, 0.05, 0.01]$&  $[4.0, 2.5, 1.0]$&   0.0058&0.0013&0.0092\\
         \hline
    \end{tabular}
    \vspace{1em}
    \caption{Mean squared errors for each experiment (noisy data). The row of $\theta$ and $k_p$ indicates the values used for each dosage (low, medium, and high dosages respectively). The noisy data was created using a noise level of 0.03. Model MSE is the MSE between the data and the model.}
    \label{tab:noisy_interp}
\end{table}

\begin{figure}[ht]
    \centering
    \subfigure[]{\includegraphics[width=0.325\textwidth]{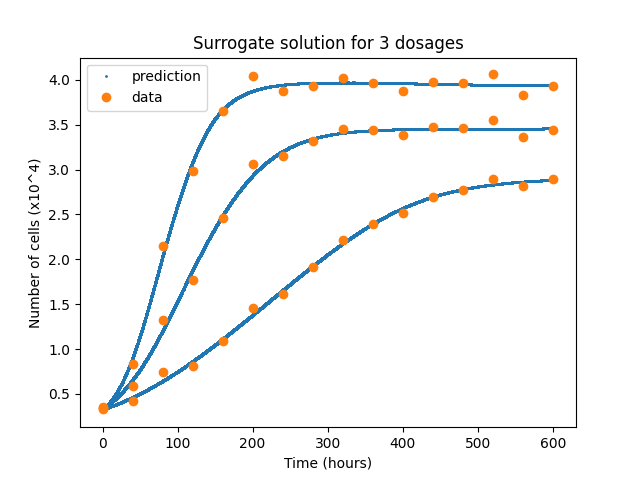}}
    \subfigure[]{\includegraphics[width=0.325\textwidth]{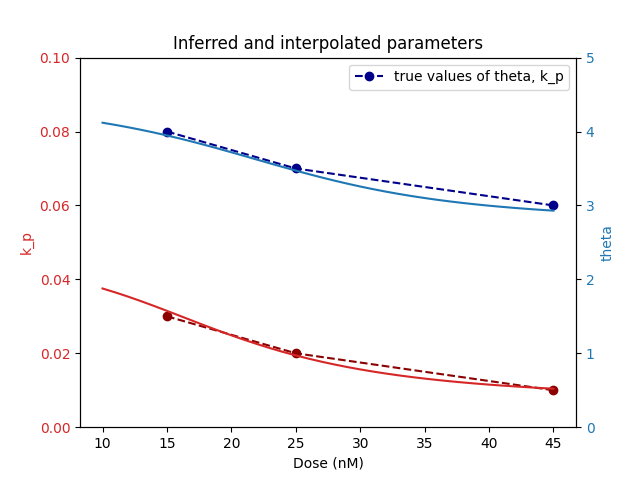}} 
    \subfigure[]{\includegraphics[width=0.325\textwidth]{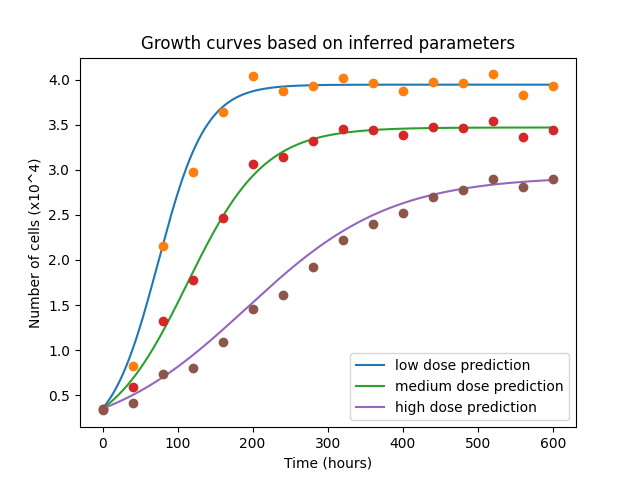}}
    \subfigure[]{\includegraphics[width=0.325\textwidth]{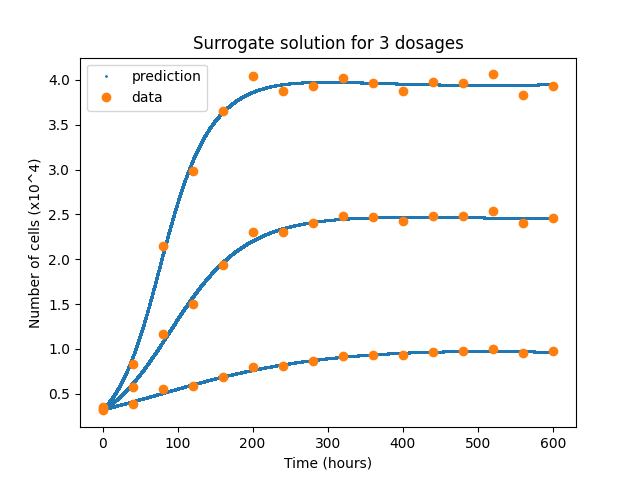}}
    \subfigure[]{\includegraphics[width=0.325\textwidth]{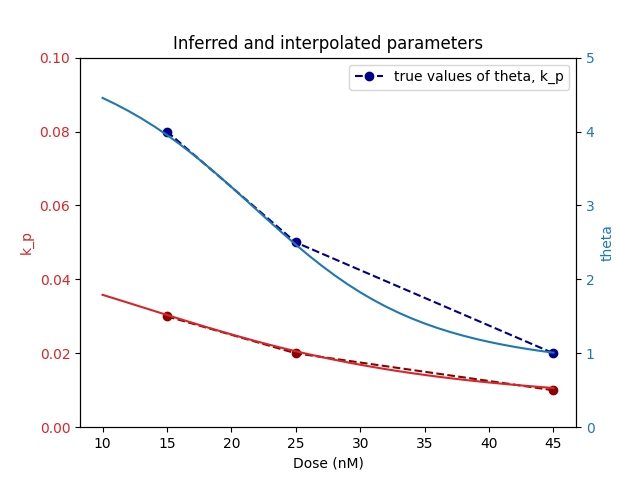}} 
    \subfigure[]{\includegraphics[width=0.325\textwidth]{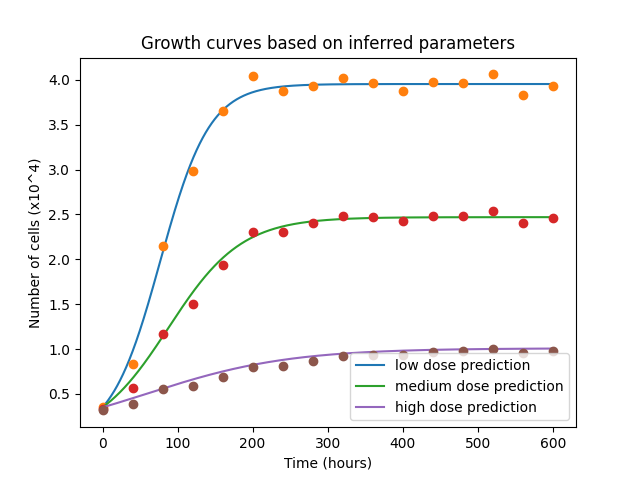}}
    \caption{(a) Surrogate solution for row 4 of Table~\ref{tab:noisy_interp}. (b) Fit and interpolated parameters per dosage. Blue is $\theta$, and red is $k_p$. (c) Data was simulated using the inferred parameters for each of the three dosages. (d)-(f) Same as (a)-(c) but for row 7 of Table~\ref{tab:noisy_interp}.}
    \label{fig:simultaneous_4_and_7}
\end{figure}

Figure~\ref{fig:simultaneous_4_and_7}, (a)-(c) shows the results for row 4 of the noisy data, and Figure~\ref{fig:simultaneous_4_and_7}, (d)-(f) shows the results for row 7 of the noisy data. It can be seen that the interpolation is performed on a smooth continuous curve, the surrogate solution is correct for each dosage, and the resulting equations, when solved using the estimated parameters, fit the data very well visually. In addition, the model MSE (the MSE between the data and the model) is on the order of $10^{-3}$. It is worth noting that sometimes, the surrogate solution returns and the fit implied by the parameters does not agree. The surrogate solution may agree with the data but the inferred parameters, when substituted into the differential equation, do not show an accurate fit. This may happen because the PINN loss is not sufficiently minimized compared to the MSE loss. For this reason, for this set of experiments, we perform both checks to ensure that the results are reasonable. Overall these results show that the UPINN method may provide an alternative to fitting multiple PINNs, with the added capability of interpolating between observations.

\newpage 

\subsection{Learning doxorubicin dynamics}

In the next set of experiments, we aim to learn doxorubicin dynamics from in-vitro data, as gathered and modelled by~\cite{mckenna2017predictive}. Due to the absence of the ground truth net proliferation rate for in-vitro experimental data, we first validate the method using simulated data. We use~\cite{mckenna2017predictive}'s ODE to generate data from eq.~\ref{eq:4} and one of~\ref{eq:5A} and~\ref{eq:5B}. We then learn $F_D(N,t)$ and $G_D(t)$ as per the methods for eq.~\ref{eq:5B}, and we learn $F_D(N,t)$ from data generated by eq.~\ref{eq:5A}. Subsequently, we learn $F_D(N,t)$ and $G_D(t)$ from the in-vitro SUM-149 data.

% The final aim of this experiment is to fit $k(t)$ from eq \ref{eq:simple_mckenna} for a sufficient number of dosages and exposure times. However, since for the experimental data we do not have ground truth for $k(t)$, we had to validate the performance of the method in a different way. We first simulated data according to equations (4) and (5) separately, using realistic parameters for each dosage. If our method can successfully recover $k(t)$ from data generated using each of theses two models, then we can be more certain it is recovering the correct drug action from the in-vitro data.

% It is worthwhile to note that while the equation

% \[\frac{dN_{TC}(t)}{dt} = (k_p - k_d(t)) N_{TC}(t) (1 -  N_{TC}(t))\]

% is also identifiable with respect to $k_d(t)$, but due to numerical approximation errors, empirically  $k_d(t)$ is not identifiable in this equation. However, it may be possible to improve the UPINN method such that it is in fact able to recover $k_d(t)$ to a high degree of accuracy.

For our in-silico analysis, we generated time series data $\{t_i,n_i\}$ (number of tumour cells over time) either via equation~\ref{eq:5A} or~\ref{eq:5B}, for different sets of realistic parameters. Since $k_{d,A}(D),k_{d,B}(D),\theta(D),r(D)$ take on scalar real values for a given dosage, it suffices to generate a dataset with one parameter combination for each ``dosage''. We generated several such datasets. By comparing it to the ground truth net proliferation rate, we are able to validate the performance of the model. Parameter values for these data were chosen to be similar to the values obtained via fitting by~\cite{mckenna2017predictive}. $k_p$ was selected to be 0.0354 $\text{hr}^{-1}$. This value was obtained by fitting a standard continuous PINN~\cite{RAISSI2019686} to the control data provided by the authors of~\cite{mckenna2017predictive}. Finally, we added noise to the synthetic data, as per synthetic experiments in~\cite{podina2022pinn} at a noise level of 0.03. The data generated for each dosage was normalized such that both the time and cell counts range from 0 to 1. In other words, the carrying capacity information and the timescale information is lost at this stage, but can be recovered later after $F_D(N,t)$ and $G_D(t)$ are fit, by undoing the scaling. This initial scaling improves the performance of the UPINN method.

% The UPINN method was set up, in all experiments in this section to learn $k(t)$ only (from eq. (6)) in addition to the surrogate solution $N_{TC}$. This is a more general form of eq (3). Hence, the hidden term learned in this formulation can then be used to estimate other terms in the model. More specifically, if we assume $\theta$ to be 1, and $k_p$ to be known from control data, then $k_d(t,D)$ can be recovered from the resulting hidden term algebraically. 

% As described in the methods, first data was simulated according to one of eq~\ref{eq:5A} and~\ref{eq:5B}, and then the UPINN method was used to learn the hidden terms. 
% 30 datapoints were simulated with noise added at a level of 0.03. 
% When simulating from eq (4), different (realistic) $k_{d,A}$ were chosen to be explored. 

\begin{figure}[ht]
    \centering
    \subfigure[]{\includegraphics[width=0.45\textwidth]{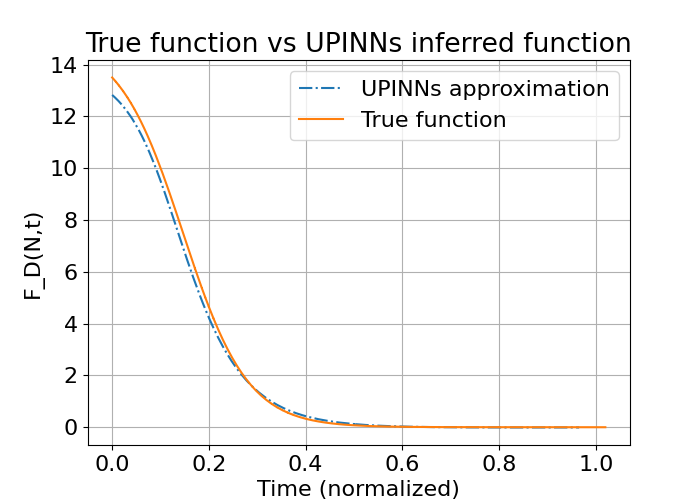}}
    \subfigure[]{\includegraphics[width=0.45\textwidth]{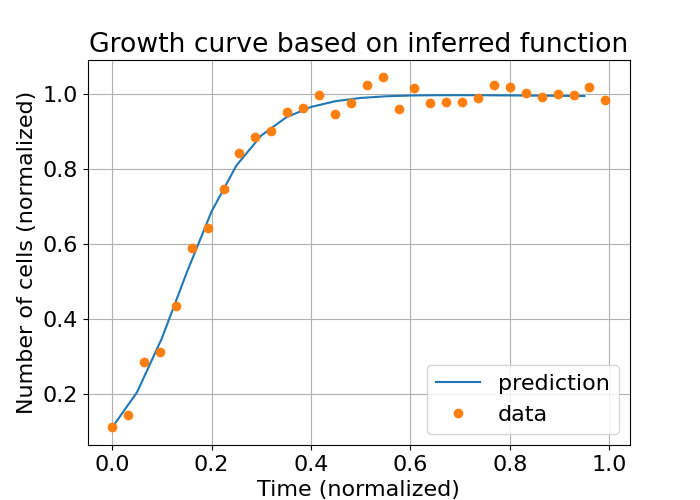}}
    \caption{(a) Fit of $F_D(N,t)$ to data from eq~\ref{eq:5A} with parameters $k_{d,A} = 0.03, k_p = 0.0354$ (b) ODE solution generated using the inferred $F_D(N,t)$, along with the data for comparison. $F_D$ is recovered well using the UPINN method, and the resulting model using the recovered $F_D$ matches the data well.}
    \label{fig:5A_F_vis}
\end{figure}

First, data from eq.~\ref{eq:4} combined with ~\ref{eq:5A} was generated, utilizing different parameter values to create datasets for different ``dosages". Table~\ref{tab:5A_res} summarizes the different parameters and noise levels tested. For the same dataset, the UPINN was used to fit $F_D(N,t)$ from eq \ref{eq:F} 5 times, and the best, mean and standard deviation MSE is reported. The MSE is computed between the predicted cell count and the actual cell count. The predicted cell count was obtained by substituting the learned $F_D(N,t)$ into eq~\ref{eq:F} and solving the equation numerically. Figure~\ref{fig:5A_F_vis} shows a particular fit of $F_D(N,t)$ using data generated from $\theta = 1.0$, $k_{d,A}=0.03$ $\text{hr}^{-1}$. The method performs well in identifying $F_D(N,t)$ for different realistic values of the parameters, as evidenced by the MSE model error being at most $10^{-4}$. Given that the standard deviation of the fits is on the order of $10^{-5}$ and $10^{-6}$, we can conclude that for this equation, the UPINN method produces reproducible fits.

% Table~\ref{tab:5A_res} summarizes the MSE of the learned $k(t)$ for several different parameters $k_{d,A}$, and noise levels. Five different experiments were run for each combination, and the $k(t)$ with the best final fit was chosen to be reported. The final fit was computed by solving the differential equation with $k(t)$ provided by the trained neural network.

 \begin{table}[ht]
     \centering
        \def\arraystretch{1.5}%
     \begin{tabular}{|c|c|c||c|c|c|} \hline 
          $\theta$&  $k_{d,A}$&  noise&  Best solution fit MSE&  Mean MSE& St. Dev. of MSE \\ \hline \hline
          1.0&  0.05&  0.0& 1.11e-05  &  4.17e-05  & 2.31e-05 \\ 
          1.0&  0.03&  0.0& 6.03e-06  & 1.42e-05    &  1.48e-05 \\
          1.0&  0.01&  0.0& 2.20e-07  & 3.24e-06 &  2.99e-06  \\  
          1.0&  0.05&  0.03& 5.83e-04 & 5.98e-04  & 2.41e-05  \\  
          1.0&  0.03&  0.03&  5.02e-04 &   5.11e-04 &  1.30e-05 \\  
          1.0&  0.01&  0.03&  2.15e-04 &  2.17e-04  &  1.71e-06 \\ \hline
     \end{tabular}
     \vspace{1em}
     \caption{Fit of $F_D(N,t)$ using data generated via eq.~\ref{eq:5A}, with MSE computed between the inferred solution (using the learned hidden term) and the data. Each experiment was run 5 times; the error of the best fit is shown, along with the mean and standard deviation of all 5 runs.}
     \label{tab:5A_res}
 \end{table}

% \begin{figure}[ht]
%     \centering
%     \subfigure[]{\includegraphics[width=0.325\textwidth]{figures/u_pred_check_26.png}}
%     \subfigure[\label{fig:noisy_2}]{\includegraphics[width=0.325\textwidth]{figures/real_vs_NN_26.png}} 
%     \subfigure[]{\includegraphics[width=0.325\textwidth]{figures/sim_data_26.png}}
%     \caption{(a) Surrogate solution and noisy data for row 5 of Table~\ref{tab:5A_res}. (b) Fit of $k(t)$ from eq \ref{eq:simple_mckenna} (c) Data simulated using the inferred function, along with the data}
%     \label{fig:simultaneous_4_and_7}
% \end{figure}

\begin{figure}[ht]
    \centering
    \subfigure[]{\includegraphics[width=0.45\textwidth]{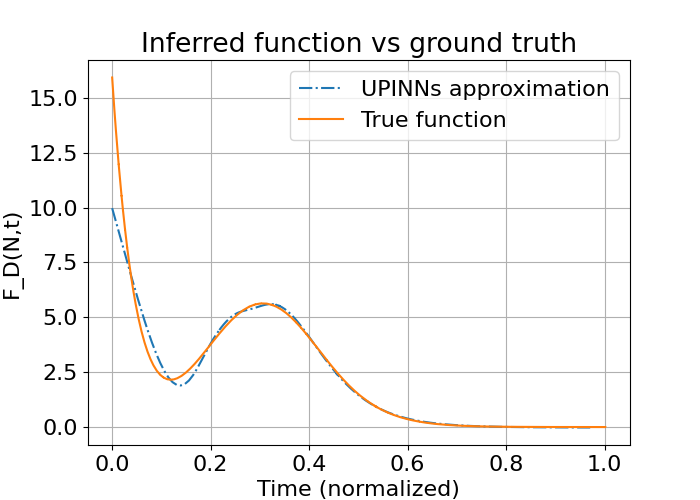}} 
    \subfigure[]{\includegraphics[width=0.45\textwidth]{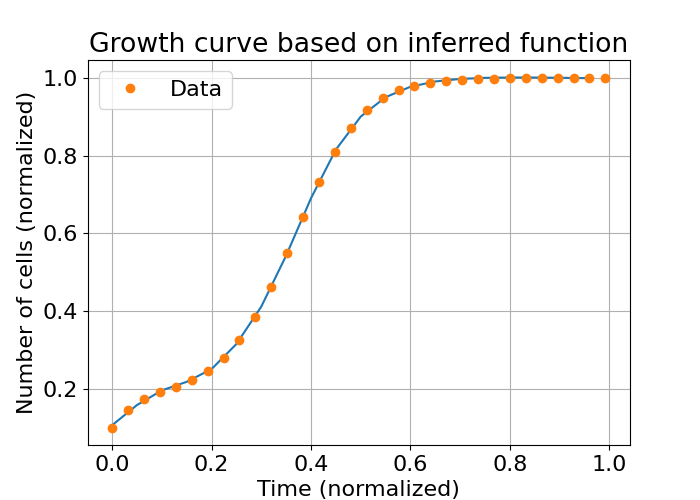}}
    \subfigure[]{\includegraphics[width=0.45\textwidth]{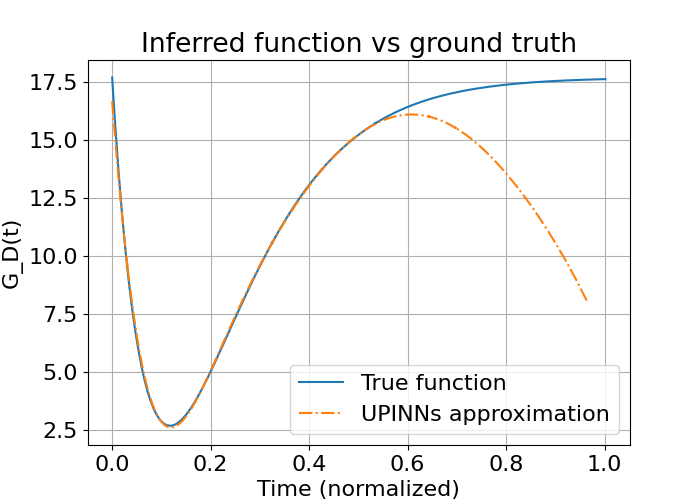}} 
    \subfigure[]{\includegraphics[width=0.45\textwidth]{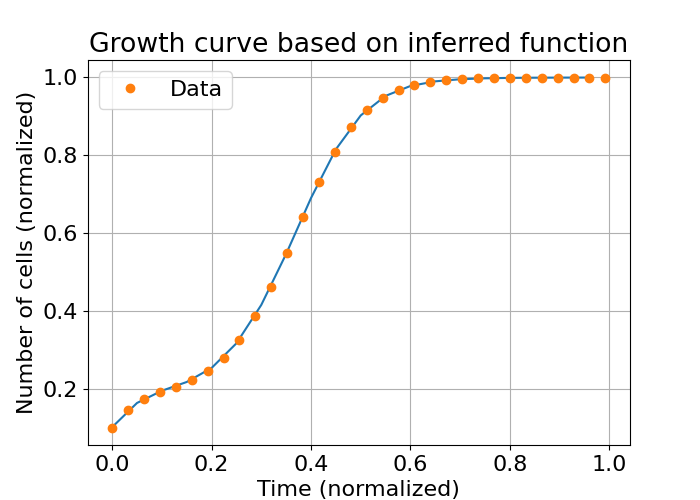}}
    \caption{\textbf{(a)} Fit of $F_D(N,t)$ (eq~\eqref{eq:F}) to data from row 2 of Table~\ref{tab:5B_res} (eq.~\ref{eq:5B}, parameters $r = 0.017, k_{d,B} = 0.03$) \textbf{(b)} ODE solution generated using the inferred $F_D(N,t)$, along with the data for comparison \textbf{(c), (d)} Same as (a), (b) but $G_D(t)$ is fit. $F_D$ in (a) is recovered with a solution MSE of $10^{-6}$, but $G_D$ in (c) appears to not be fully identifiable from the data. }
    %(a) The numerically-solved solution to the ODE using the inferred function. Dataset taken from row 2 of Table~\ref{tab:5B_res} (equations~\ref{eq:4} and~\ref{eq:5B}) (b) The inferred function $F_D(N,t)$ for the same dataset}
    \label{fig:5B_F_G_vis}
\end{figure}

Next, data from eq.~\ref{eq:4} combined with~\ref{eq:5B} was generated. In this case, the net proliferation rate is time-dependent, whereas in eq.~\ref{eq:5A} it is constant. Table \ref{tab:5B_res} summarizes the MSE of the learned $F_D(N,t)$ when the data was generated according to eq~\ref{eq:5B}. We can see that the model MSE is on the order of at most $10^{-4}$ for most parameter combinations. In this case, several different $(r_d,k_{d,B})$ combinations are chosen, and the best runs of 5 are chosen similarly. Since the standard deviation is at most $10^{-3}$ (unless one of the runs completely fails to learn the underlying function), we do not anticipate that adding more runs would change these MSEs significantly. The values chosen for these parameters were informed in part by~\cite{mckenna2017predictive} such that a variety of realistic growth curves were generated. Carrying capacity $\theta$ is set to 1 to avoid the additional step of rescaling the cell counts. Table~\ref{tab:5B_res} reports the results for both noisy and noiseless data. It can be seen that the very first row had a run that has an error on the order of $10^1$, but the MSE for the best model run is still $10^{-5}$. This means that for this particular parameter combination, many of the model fits were not accurate, but due to the stochasticity of the neural network fitting process, it is still possible to obtain a model with MSE less than $10^{-3}$ within 5 trials. Figure~\ref{fig:5B_F_G_vis} (a), (b) shows a fit of $F_D(N,t)$ to row 2 of Table~\ref{tab:5B_res}. The solution shows a time-dependent net proliferation rate, which is learned with a model MSE of $10^{-6}$.  We also demonstrate that $G_D(t)$ can be learned well from equation~\eqref{eq:5B}. Figure~\ref{fig:5B_F_G_vis} (c), (d) shows the fit of the same row 2, with parameters $r_d=0.017$ and $k_{d,B}=0.05$, but this time fitting $G_D(t)$. In this case, the solution matches the ground truth very closely except for $t \in [0.6,1.0]$. At this point, the predictions start to diverge. This is likely because the function ceases to be identifiable in this time interval.
% Especially from noiseless data, the net proliferation rate can be learned such that the model MSE is $10^{-6}$.

 \begin{table}[ht]
     \centering
         \def\arraystretch{1.5}%
     \begin{tabular}{|c|c|c|c|c|c|c|} \hline 
          $r_{d}$& $\theta$&  $k_{d,B}$&  noise&  Best solution MSE&  Mean MSE& St. Dev. of MSE \\ \hline \hline
          0.017 & 1.0&  0.05&  0.0& 1.83e-05  &  23.0  & 46.0 \\ 
          0.017 & 1.0&  0.03&  0.0& 4.38e-06  & 8.06e-05    &  6.14e-05 \\ 
          0.017 & 1.0&  0.01&  0.0& 1.35e-05  & 2.59e-05 &  9.87e-06   \\ 
          0.017 & 1.0&  0.05&  0.03& 2.28e-04 & 1.03e-02 & 6.56e-03  \\  
          0.017 & 1.0&  0.03&  0.03&  4.37e-04 &   4.99e-04 & 7.25e-05 \\  
          0.017 & 1.0&  0.01&  0.03&  5.00e-04 &  5.11e-04 &  8.82e-06 \\ \hline
     \end{tabular}
     \vspace{1em}
     \caption{$F_D(N,t)$ (eq~\ref{eq:F}) using data generated via eq.~\ref{eq:5B}, with MSE computed between the inferred solution (using the learned hidden term) and the data. Each experiment was run 5 times; the error of the best fit is shown, along with the mean and standard deviation of all 5 runs.}
     \label{tab:5B_res}
 \end{table}

%  \begin{figure}[ht]
%     \centering
%     \subfigure[]{\includegraphics[width=0.45\textwidth]{figures/5B_G_soln.png}}
%     \subfigure[]{\includegraphics[width=0.45\textwidth]{figures/5B_G_inferred.png}} 
%     \caption{(a) The numerically-solved solution to the ODE using inferred $F_D(N,t)$. Dataset taken from row 2 of Table~\ref{tab:5B_res} (equation~\ref{eq:4} and~\ref{eq:5B}) (b) The inferred function $F_D(N,t)$ for the same dataset}
%     \label{fig:5B_F_vis}
% \end{figure}

% Similarly, Figure \ref{fig:5B_stuff} shows the final fits of rows 1 and 2 of Table 5. Both the surrogate solution and the numerical solution fit the noisy data very well. However, the method fails to learn the true function much of the time, under noisy conditions. For this reason, certain runs significantly increased the mean and standard deviation of the runs in row 1 of this table.

\begin{figure}[ht]
    \centering
    \subfigure[]{\includegraphics[width=0.45\textwidth]{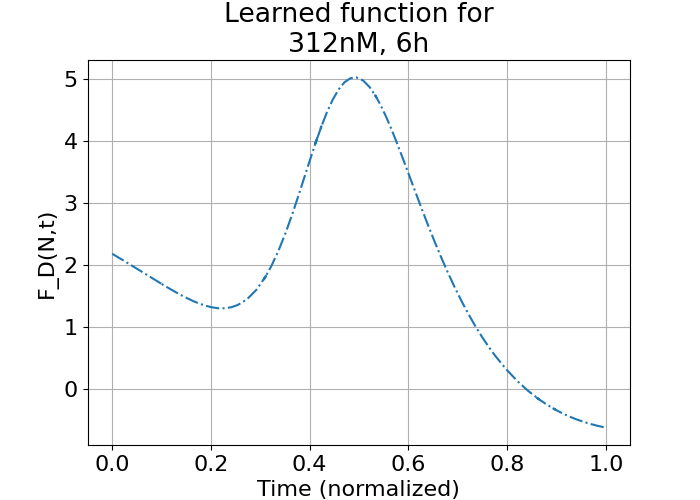}} 
    \subfigure[]{\includegraphics[width=0.45\textwidth]{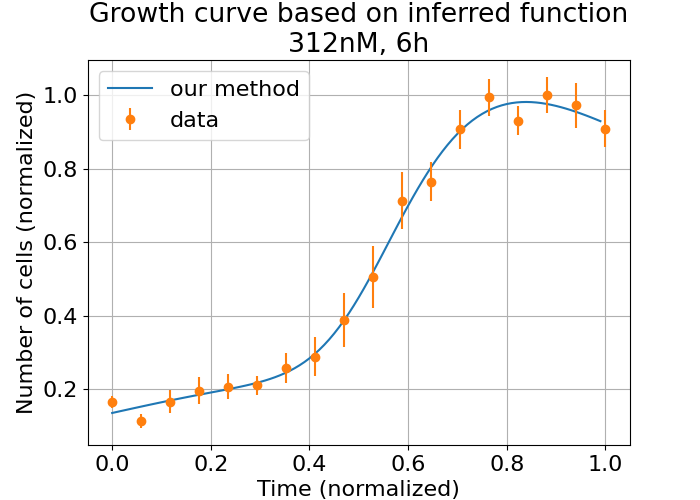}}
    \subfigure[]{\includegraphics[width=0.45\textwidth]{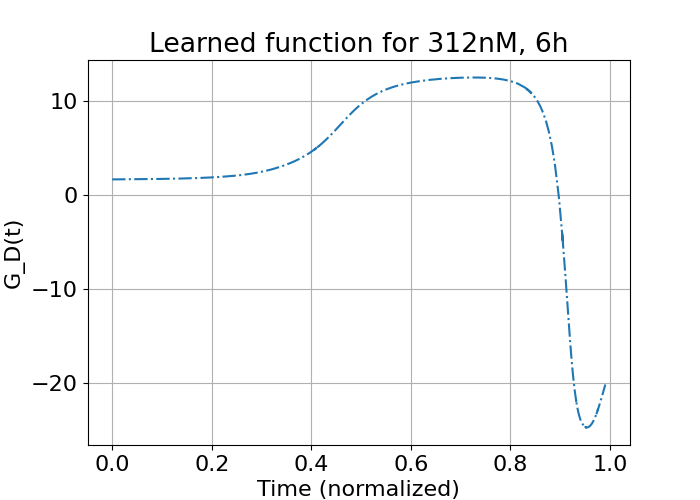}}
    \subfigure[]{\includegraphics[width=0.45\textwidth]{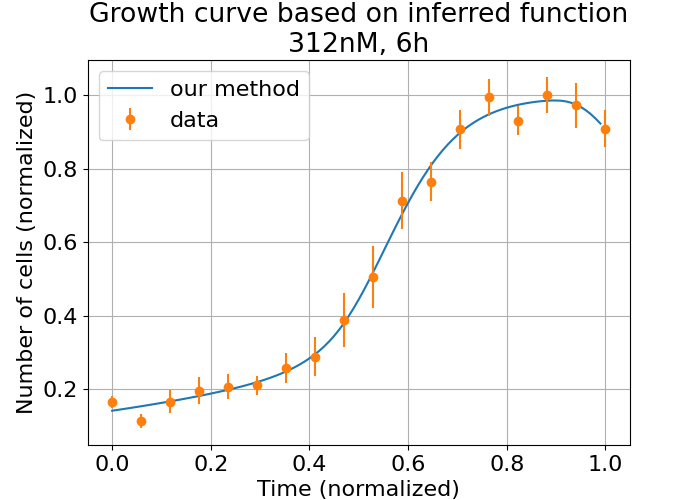}}
    \caption{\textbf{(a)} Fit of $F_D(N,t)$ (eq~\eqref{eq:F}) to experimental data: 312nM, 6h exposure \textbf{(b)} ODE solution generated using the inferred $F_D(N,t)$, along with the data for comparison \textbf{(c), (d)} Same as (a), (b) but $G_D(t)$ (eq~\eqref{eq:G}) is fit. The overall curve of (a) is similar in shape visually to Fig. 9 (a), indicating that the growth is captured to some extent by Model~\eqref{eq:4}. Additionally, (c) shows a plateau in the function before an increase in cell killing, which cannot be captured by a linear combination of~\eqref{eq:5A} and~\eqref{eq:5B}.}
    % \caption{(a) The numerically-solved solution to the ODE using the inferred function. Dataset taken from row 2 of Table~\ref{tab:5B_res} (b) The inferred function $F_D(N,t)$ for the same dataset}
    \label{fig:real_F_G_vis}
\end{figure}

Since the UPINN method is able to accurately recover the hidden terms $F_D(N,t)$ and $G_D(t)$ for both model~\ref{eq:5A} and~\ref{eq:5B}, we apply the UPINN method to learn $F_D(N,t)$ and $G_D(t)$ for in-vitro data, where no ground truth is available. Figure~\ref{fig:real_F_G_vis} shows the learned time-dependent term $F_D(N,t)$ for the application of the doxorubicin chemotherapeutic to SUM-149 cells, at 312nM concentration, for an exposure time of 6hrs. The cells were then allowed to grow undisturbed. This curve is very similar in shape to the one shown in Fig.~\ref{fig:5B_F_G_vis} (a). The peak occurs at approximately the same height as well. Figure~\ref{fig:real_F_G_vis} (c), (d) shows the time-dependent net proliferation, $G_D(t)$, learned for the same in-vitro dataset. We can see that this curve is not possible to construct using equation~\ref{eq:5A} or~\ref{eq:5B} alone, or even with a weighted average of these equations. This is because the function shown is very flat up until $t=0.4$, but the derivative increases after this point. By contrast, a weighted average of~\ref{eq:5A} and~\ref{eq:5B} could not have a derivative of zero for any finite time interval given their formulations. Hence, fitting the UPINN to find $G_D(t)$ provides insights into the true net proliferation rate as a function of time. Initially, the net proliferation rate behaves more like~\ref{eq:5A}, and subsequently more like~\ref{eq:5B}. Table \ref{tab:mckenna_res} shows the MSE between the solution using the learned function $F_D(N,t)$ and the data. The inferred solution fits the data well in most cases, as evidenced by the MSE between the inferred solution and the data being on the order of $10^{-4}$ for over half of the time-series datasets.

\section{Discussion}

In this work, we apply the UPINN method to learn the chemotherapy drug action in several different ODEs, applying the method to both synthetic and in-vitro experimental data. Rather than making assumptions about the drug action, we can learn it from data in order to identify how to best model the effect of the chemotherapeutic on cells over time. In addition, this learned drug action may allow us to learn more about the underlying biological mechanism. Finally, the learned drug action can then be used for downstream tasks such as drug treatment schedule optimization. Current limitations of this method is a lack of uncertainty quantification, which is especially important given the method's stochastic nature and potential high-risk downstream applications, lack of integration with model identifiability workflows, and minimal quantification of differential equation setups where it is unlikely to perform well.

% Limitations of this work include the lack of uncertainty quantification, lack of integration with model identifiability assessments. Additionally, it is yet unknown under which conditions the hidden component can be learned, and when not. Uncertainty quantification has been proposed for PINNs~\cite{yang2021b}, but it would be especially help to quantify the uncertainty for UPINNs. Additionally, it would be beneficial if the uncertainty quantification method has validity guarantees, such as conformal prediction~\cite{}, and if it was integrated with existing literature about structural and practical identifiability of differential equations~\cite{}. This could potentially enable the application of the learned drug action in a higher-stakes scenario. Finally, more investigation is needed on what kind of functions or relationships the method can learn and which it cannot. Neural networks can sometimes fail at modelling highly oscillatory behaviour~\cite{}, so quantifying the properties of functions that are easy or hard for UPINNs to learn is important for downstream tasks.

% \section{Future Work}

General future work in this direction would involve understanding the conditions under which this method performs well and when it is not able to learn the drug action effectively (e.g. noise level, shape and properties of the ground truth function). If there is more than one function to be learned, understanding the identifiability of the functions would provide more guidance on whether the method can successfully recover the ground truth. This could be a mathematical analysis similar to parameter identifiability, or it could be a more nuanced quantification of the uncertainty of the output of the neural networks, taking into account the abundance of data. Uncertainty quantification has been proposed for PINNs~\cite{yang2021b}, but it would be especially helpful to quantify the uncertainty for UPINNs. Additionally, it would be beneficial if the uncertainty quantification method has validity guarantees, such as conformal prediction~\cite{angelopoulos2021gentle}. Finally, applying the method to a drug with a known model of drug induced death (e.g. Norton-Simon) would further validate the method's performance. As an additional step after the functions have been learned, symbolic regression could be performed to find a closed form of the function.

As an extension to learning the dynamics of doxorubicin, the UPINN method can be easily adapted to learn the unknown net proliferation rate $G(t,D)$ as a function of dosage and time (and possibly cells) simultaneously. This would involve treating the number of cells as a function of both dosage and time as well. However, there are several considerations: due to the limited dosage data (only 9 measurements), interpolation between these measurements may not yield good results. However, it should be possible to fit $G(t,D)$ correctly for the available dosages. Secondly, there may be identifiability issues and there may not be a single unique $G$ which satisfies the equation.

\section{Conclusion}

In this paper, we integrate machine learning in the form of the Universal Physics-Informed Neural Network (UPINN) method with QSP models in order to learn the drug action of chemotherapeutics accurately and with minimal computational expenses. We showcase the ability of the method to identify three different well-known drug actions, the Log-kill model, Norton-Simon, and $E_{max}$. The learned drug actions match the ground truth very well. This method can also be used to infer many parameter sets simultaneously rather than running a separate fitting procedure for each dataset. In addition, the method can interpolate between fitted parameters. Finally, we employ the method to learn the drug action of doxorubicin from time-series data.

\section{Acknowledgements}

We thank Matthew McKenna (Vanderbilt University), Thomas Yankeelov (The University of Texas at Austin), and Ernesto Lima (The University of Texas at Austin) for sharing their data and for informative discussions and comments on the first draft of the manuscript. We acknowledge the support of the Natural Sciences and Engineering Research Council of Canada (NSERC).

% \begin{figure}[ht]
%     \centering
%     \subfigure[]{\includegraphics[width=0.3\textwidth]{figures/u_pred_check_13.png}}
%     \subfigure[]{\includegraphics[width=0.3\textwidth]{figures/real_vs_NN_13.png}} 
%     \subfigure[]{\includegraphics[width=0.3\textwidth]{figures/sim_data_13.png}}
%     \subfigure[]{\includegraphics[width=0.3\textwidth]{figures/u_pred_check_21.png}}
%     \subfigure[]{\includegraphics[width=0.3\textwidth]{figures/real_vs_NN_21.png}} 
%     \subfigure[]{\includegraphics[width=0.3\textwidth]{figures/sim_data_21.png}}
%     \caption{(a) The surrogate solution of row 1 of Table 5 (b) The inferred function for the same dataset (c) The numerically-solved solution to the ODE using the inferred function, same dataset. (d), (e), (f): The same plots, but for row 2 of Table 5.}
%     \label{fig:5B_stuff}
% \end{figure}

% Having established that the method performs well on simulated data, we now test it on real data. We perform the same fit (using UPINNs to learn $k(t)$ from eq \ref{eq:simple_mckenna}) to all of the SUM-149 cell line data, which includes different dosages and exposure times. The results from two different fits can be seen in Fig \ref{fig:in_vitro}. 

 \begin{table}[ht]
     \centering
         \def\arraystretch{1.5}%
     \begin{tabular}{|c|c|c|c|c|} 
     \hline 
     Duration & Concentration &  Best MSE & Mean MSE & St. Dev. of MSE\\ \hline \hline
        6h & 10nM & 0.00045 &  0.0022 & 0.0036 \\
        6h & 20nM & 0.00031 &  0.0011 & 0.0013 \\
        6h & 39nM & 7.57e-05 &  0.024 & 0.062 \\
        6h & 78nM & 0.00050 &  0.0013 & 0.00054 \\
        6h & 156nM & 0.00082 &  0.00084 & 8.04e-06 \\
        6h & 312nM & 0.00019 &  0.00086 & 0.00049 \\
        6h & 625nM & 0.00061 &  0.010 & 0.016 \\
        6h & 1250nM & 0.00038 &  0.0072 & 0.0065 \\
        6h & 2500nM & 0.0047 &  0.0062 & 0.00097 \\
        \hline
        12h & 10nM & 0.00022 &  0.0011 & 0.0015 \\
        12h & 20nM & 0.00066 &  0.0021 & 0.0015 \\
        12h & 39nM & 0.00056 &  0.0011 & 0.00020 \\
        12h & 78nM & 0.00084 &  0.00086 & 7.50e-06 \\
        12h & 156nM & 0.00079 &  0.0021 & 0.0032 \\
        12h & 312nM & 0.00011 &  0.021 & 0.036 \\
        12h & 625nM & 0.00043 &  0.012 & 0.033 \\
        12h & 1250nM & 0.00086 &  0.0076 & 0.0031 \\
        12h & 2500nM & 0.0014 &  0.011 & 0.0098 \\
        \hline
        24h & 10nM & 0.00050 &  0.00065 & 0.00013 \\
        24h & 20nM & 0.00067 &  0.0014 & 0.00097 \\
        24h & 39nM & 0.00092 &  0.0011 & 0.00010 \\
        24h & 78nM & 0.00043 &  0.00098 & 0.00024 \\
        24h & 156nM & 0.00062 &  0.011 & 0.019 \\
        24h & 312nM & 0.00031 &  0.0023 & 0.0040 \\
        24h & 625nM & 0.00033 &  0.0063 & 0.0038 \\
        24h & 1250nM & 0.0048 &  0.0048 & 6.66e-05 \\
        24h & 2500nM & 0.0045 &  0.0078 & 0.0013 \\ \hline
     \end{tabular}
     \vspace{1em}
     \caption{Fits of $F_D(N,t)$ to in-vitro SUM-149 time-series data from~\cite{mckenna2017predictive}. MSE is computed between the inferred solution (using the learned hidden term) and the in-vitro data. Each experiment was run 10 times; the error of the best run is shown, along with the mean and standard deviation of all 10 runs.}
     \label{tab:mckenna_res}
 \end{table}

\bibliographystyle{plain.bst}
\bibliography{references}  %%% Uncomment this line and comment out the ``thebibliography'' section below to use the external .bib file (using bibtex) .

\begin{thebibliography}{10}

\bibitem{allen2016efficient}
RJ~Allen, Theodore~R Rieger, and Cynthia~J Musante.
\newblock Efficient generation and selection of virtual populations in quantitative systems pharmacology models.
\newblock {\em CPT: pharmacometrics \& systems pharmacology}, 5(3):140--146, 2016.

\bibitem{angelopoulos2021gentle}
Anastasios~N Angelopoulos and Stephen Bates.
\newblock A gentle introduction to conformal prediction and distribution-free uncertainty quantification.
\newblock {\em arXiv preprint arXiv:2107.07511}, 2021.

\bibitem{basse2004modelling}
Britta Basse, Bruce~C Baguley, Elaine~S Marshall, Wayne~R Joseph, Bruce van Brunt, Graeme Wake, and David~JN Wall.
\newblock Modelling cell death in human tumour cell lines exposed to the anticancer drug paclitaxel.
\newblock {\em Journal of Mathematical Biology}, 49:329--357, 2004.

\bibitem{bolton2015proposed}
Larisse Bolton, Alain~HJJ Cloot, Schalk~W Schoombie, and Jacobus~P Slabbert.
\newblock A proposed fractional-order gompertz model and its application to tumour growth data.
\newblock {\em Mathematical medicine and biology: a journal of the IMA}, 32(2):187--209, 2015.

\bibitem{camacho2018next}
Diogo~M Camacho, Katherine~M Collins, Rani~K Powers, James~C Costello, and James~J Collins.
\newblock Next-generation machine learning for biological networks.
\newblock {\em Cell}, 173(7):1581--1592, 2018.

\bibitem{derbalah2022framework}
Abdallah Derbalah, Hesham Al-Sallami, Chihiro Hasegawa, Abhishek Gulati, and Stephen~B Duffull.
\newblock A framework for simplification of quantitative systems pharmacology models in clinical pharmacology.
\newblock {\em British Journal of Clinical Pharmacology}, 88(4):1430--1440, 2022.

\bibitem{drexler2020experimental}
D{\'a}niel~Andr{\'a}s Drexler, Tam{\'a}s Ferenci, Andr{\'a}s F{\"u}redi, Gergely Szak{\'a}cs, and Levente Kov{\'a}cs.
\newblock Experimental data-driven tumor modeling for chemotherapy.
\newblock {\em IFAC-PapersOnLine}, 53(2):16245--16250, 2020.

\bibitem{eastman2021reinforcement}
Brydon Eastman, Michelle Przedborski, and Mohammad Kohandel.
\newblock Reinforcement learning derived chemotherapeutic schedules for robust patient-specific therapy.
\newblock {\em Scientific Reports}, 11(1):17882, 2021.

\bibitem{eduati2020patient}
Federica Eduati, Patricia Jaaks, Jessica Wappler, Thorsten Cramer, Christoph~A Merten, Mathew~J Garnett, and Julio Saez-Rodriguez.
\newblock Patient-specific logic models of signaling pathways from screenings on cancer biopsies to prioritize personalized combination therapies.
\newblock {\em Molecular systems biology}, 16(2):e8664, 2020.

\bibitem{jagtap2020extended}
Ameya~D Jagtap and George~Em Karniadakis.
\newblock Extended physics-informed neural networks (xpinns): A generalized space-time domain decomposition based deep learning framework for nonlinear partial differential equations.
\newblock {\em Communications in Computational Physics}, 28(5), 2020.

\bibitem{jarrett2019experimentally}
Angela~M Jarrett, Alay Shah, Meghan~J Bloom, Matthew~T McKenna, David~A Hormuth, Thomas~E Yankeelov, and Anna~G Sorace.
\newblock Experimentally-driven mathematical modeling to improve combination targeted and cytotoxic therapy for her2+ breast cancer.
\newblock {\em Scientific reports}, 9(1):12830, 2019.

\bibitem{kingma2014adam}
Diederik~P Kingma and Jimmy Ba.
\newblock Adam: A method for stochastic optimization.
\newblock {\em arXiv preprint arXiv:1412.6980}, 2014.

\bibitem{kohandel2006mathematical}
M~Kohandel, S~Sivaloganathan, and A~Oza.
\newblock Mathematical modeling of ovarian cancer treatments: sequencing of surgery and chemotherapy.
\newblock {\em Journal of theoretical biology}, 242(1):62--68, 2006.

\bibitem{liu1989limited}
Dong~C Liu and Jorge Nocedal.
\newblock On the limited memory bfgs method for large scale optimization.
\newblock {\em Mathematical programming}, 45(1-3):503--528, 1989.

\bibitem{mckenna2017predictive}
Matthew~T McKenna, Jared~A Weis, Stephanie~L Barnes, Darren~R Tyson, Michael~I Miga, Vito Quaranta, and Thomas~E Yankeelov.
\newblock A predictive mathematical modeling approach for the study of doxorubicin treatment in triple negative breast cancer.
\newblock {\em Scientific reports}, 7(1):5725, 2017.

\bibitem{panetta2003optimal}
John~Carl Panetta and K~Renee Fister.
\newblock Optimal control applied to competing chemotherapeutic cell-kill strategies.
\newblock {\em SIAM Journal on Applied Mathematics}, 63(6):1954--1971, 2003.

\bibitem{paszke2019pytorch}
Adam Paszke, Sam Gross, Francisco Massa, Adam Lerer, James Bradbury, Gregory Chanan, Trevor Killeen, Zeming Lin, Natalia Gimelshein, Luca Antiga, et~al.
\newblock Pytorch: An imperative style, high-performance deep learning library.
\newblock {\em Advances in neural information processing systems}, 32, 2019.

\bibitem{podina2022pinn}
Lena Podina, Brydon Eastman, and Mohammad Kohandel.
\newblock A pinn approach to symbolic differential operator discovery with sparse data.
\newblock {\em arXiv preprint arXiv:2212.04630}, 2022.

\bibitem{polak2019better}
Sebastian Polak, Zofia Tylutki, Mark Holbrook, and Barbara Wi{\'s}niowska.
\newblock Better prediction of the local concentration--effect relationship: the role of physiologically based pharmacokinetics and quantitative systems pharmacology and toxicology in the evolution of model-informed drug discovery and development.
\newblock {\em Drug discovery today}, 24(7):1344--1354, 2019.

\bibitem{przedborski2021systems}
Michelle Przedborski, Munisha Smalley, Saravanan Thiyagarajan, Aaron Goldman, and Mohammad Kohandel.
\newblock Systems biology informed neural networks (sbinn) predict response and novel combinations for pd-1 checkpoint blockade.
\newblock {\em Communications Biology}, 4(1):877, 2021.

\bibitem{puniya2021integrative}
Bhanwar~Lal Puniya, Rada Amin, Bailee Lichter, Robert Moore, Alex Ciurej, Sydney~J Bennett, Ab~Rauf Shah, Matteo Barberis, and Tom{\'a}{\v{s}} Helikar.
\newblock Integrative computational approach identifies drug targets in cd4+ t-cell-mediated immune disorders.
\newblock {\em NPJ systems biology and applications}, 7(1):4, 2021.

\bibitem{rackauckas2020universal}
Christopher Rackauckas, Yingbo Ma, Julius Martensen, Collin Warner, Kirill Zubov, Rohit Supekar, Dominic Skinner, Ali Ramadhan, and Alan Edelman.
\newblock Universal differential equations for scientific machine learning.
\newblock {\em arXiv preprint arXiv:2001.04385}, 2020.

\bibitem{RAISSI2019686}
M.~Raissi, P.~Perdikaris, and G.E. Karniadakis.
\newblock Physics-informed neural networks: A deep learning framework for solving forward and inverse problems involving nonlinear partial differential equations.
\newblock {\em Journal of Computational Physics}, 378:686--707, 2019.

\bibitem{sorger2011quantitative}
Peter~K Sorger, Sandra~RB Allerheiligen, Darrell~R Abernethy, Russ~B Altman, Kim~LR Brouwer, Andrea Califano, David~Z D’Argenio, Ravi Iyengar, William~J Jusko, Richard Lalonde, et~al.
\newblock Quantitative and systems pharmacology in the post-genomic era: new approaches to discovering drugs and understanding therapeutic mechanisms.
\newblock In {\em An NIH white paper by the QSP workshop group}, volume~48, pages 1--47. NIH Bethesda Bethesda, MD, 2011.

\bibitem{traina2010optimizing}
Tiffany~A Traina, Ute Dugan, Brian Higgins, Kenneth Kolinsky, Maria Theodoulou, Clifford~A Hudis, and Larry Norton.
\newblock Optimizing chemotherapy dose and schedule by norton-simon mathematical modeling.
\newblock {\em Breast disease}, 31(1):7--18, 2010.

\bibitem{udrescu2020ai}
Silviu-Marian Udrescu and Max Tegmark.
\newblock Ai feynman: A physics-inspired method for symbolic regression.
\newblock {\em Science Advances}, 6(16):eaay2631, 2020.

\bibitem{wang2019model}
Yaning Wang, Hao Zhu, Rajanikanth Madabushi, Qi~Liu, Shiew-Mei Huang, and Issam Zineh.
\newblock Model-informed drug development: current us regulatory practice and future considerations.
\newblock {\em Clinical Pharmacology \& Therapeutics}, 105(4):899--911, 2019.

\bibitem{wong2019estimation}
Chi~Heem Wong, Kien~Wei Siah, and Andrew~W Lo.
\newblock Estimation of clinical trial success rates and related parameters.
\newblock {\em Biostatistics}, 20(2):273--286, 2019.

\bibitem{yang2021b}
Liu Yang, Xuhui Meng, and George~Em Karniadakis.
\newblock B-pinns: Bayesian physics-informed neural networks for forward and inverse pde problems with noisy data.
\newblock {\em Journal of Computational Physics}, 425:109913, 2021.

\bibitem{twoheads}
Tongli Zhang, Ioannis~P Androulakis, Peter Bonate, Limei Cheng, Tom{\'a}{\v{s}} Helikar, Jaimit Parikh, Christopher Rackauckas, Kalyanasundaram Subramanian, and Carolyn~R Cho.
\newblock Two heads are better than one: current landscape of integrating qsp and machine learning: an isop qsp sig white paper by the working group on the integration of quantitative systems pharmacology and machine learning.
\newblock {\em Journal of Pharmacokinetics and Pharmacodynamics}, 49(1):5--18, 2022.

\end{thebibliography}





% \end{thebibliography}

\end{document}